\documentclass[sigconf,natbib=true]{acmart}

\usepackage{algorithm}
\usepackage{algorithmic}
\usepackage{lscape}

\usepackage{tabularx}
\usepackage{multirow}
\usepackage{float}
\usepackage[normalem]{ulem}
\usepackage{afterpage}

\usepackage{caption}
\usepackage{subfigure}
\usepackage[T1]{fontenc}
\usepackage{subcaption}
\usepackage{amsmath,amsfonts}
\usepackage{algorithmic}
\usepackage{graphicx}
\usepackage{textcomp}
\usepackage[dvipsnames]{xcolor}
\usepackage{color, colortbl}
\definecolor{Gray}{gray}{0.9}
\usepackage{amsmath}
\usepackage{pifont}

\usepackage{makecell}
\usepackage{wrapfig}
\usepackage{picinpar}
\usepackage{subcaption}
\usepackage[utf8]{inputenc} 
\usepackage{url}            
\usepackage{booktabs}       
\usepackage{amsfonts}       
\usepackage{nicefrac}       
\usepackage{microtype}      

\usepackage{algorithmic}
\usepackage{graphicx}
\usepackage{textcomp}
\usepackage{xcolor}
\usepackage{multirow}
\usepackage{algorithm}
\usepackage{bm}
\usepackage{subfigure}
\usepackage{caption}

\usepackage{CJKutf8}
\usepackage[T1]{fontenc}
\definecolor{mydarkred}{rgb}{0.6,0,0}

\newcommand{\MYhref}[3][mydarkred]{\href{#2}{\color{#1}{#3}}}

\usepackage{enumitem}

\AtBeginDocument{%
  \providecommand\BibTeX{{%
    \normalfont B\kern-0.5em{\scshape i\kern-0.25em b}\kern-0.8em\TeX}}}

\setcopyright{acmlicensed}
\copyrightyear{2018}
\acmYear{2018}
\acmDOI{XXXXXXX.XXXXXXX}

\acmConference[Conference acronym 'XX]{Make sure to enter the correct
  conference title from your rights confirmation email}{June 03--05,
  2018}{Woodstock, NY}

\acmISBN{978-1-4503-XXXX-X/18/06}

\copyrightyear{2025}
\acmYear{2025}
\setcopyright{cc}
\setcctype{by}
\acmConference[SIGIR-AP 2025]{Proceedings of the 2025 Annual International
ACM SIGIR Conference on Research and Development in Information Retrieval in
the Asia Pacific Region}{December 7--10, 2025}{Xi'an, China}
\acmBooktitle{Proceedings of the 2025 Annual International ACM SIGIR
Conference on Research and Development in Information Retrieval in the Asia
Pacific Region (SIGIR-AP 2025), December 7--10, 2025, Xi'an,
China}\acmDOI{10.1145/3767695.3769481}
\acmISBN{979-8-4007-2218-9/2025/12}

\begin{document}

\title[MARCO]{MARCO: A Cooperative Knowledge Transfer Framework for Personalized Cross-domain Recommendations}

\author{Lili Xie}
\authornote{Both authors contributed equally to this work.}
\email{xaiverlilix@gmail.com}
\affiliation{\institution{Independent Researcher}
  \city{Shen Zhen}
  \country{China}}
\author{Yi Zhang}
\authornotemark[1]
\email{uqyzha91@uq.edu.au}
\affiliation{\institution{The University of Queensland \\ CSIRO DATA61}
  \city{Brisbane}
  \country{Australia}}
\author{Ruihong Qiu}
\email{r.qiu@uq.edu.au}
\affiliation{\institution{The University of Queensland}
  \city{Brisbane}
  \country{Australia}}
\author{Jiajun Liu}
\email{jiajun.liu@csiro.au}
\affiliation{\institution{CSIRO DATA61 \\ The University of Queensland}
  \city{Brisbane}
  \country{Australia}}
\author{Sen Wang}
\email{sen.wang@uq.edu.au}
\affiliation{\institution{The University of Queensland}
  \city{Brisbane}
  \country{Australia}}

\renewcommand{\shortauthors}{Lili Xie, Yi Zhang, Ruihong Qiu, Jiajun Liu, \& Sen Wang}

\begin{abstract}
Recommender systems frequently encounter data sparsity issues, particularly when addressing cold-start scenarios involving new users or items. Multi-source cross-domain recommendation (CDR) addresses these challenges by transferring valuable knowledge from multiple source domains to enhance recommendations in a target domain. However, existing reinforcement learning (RL)-based CDR methods typically rely on a single-agent framework, leading to negative transfer issues caused by inconsistent domain contributions and inherent distributional discrepancies among source domains. To overcome these limitations, MARCO, a Multi-Agent Reinforcement Learning-based Cross-Domain recommendation framework, is proposed. It leverages cooperative multi-agent reinforcement learning, where each agent is dedicated to estimating the contribution from an individual source domain, effectively managing credit assignment and mitigating negative transfer. In addition, an entropy-based action diversity penalty is introduced to enhance policy expressiveness and stabilize training by encouraging diverse agents' joint actions. Extensive experiments across four benchmark datasets demonstrate MARCO's superior performance over state-of-the-art methods, highlighting its robustness and strong generalization capabilities. The code is at \MYhref{https://github.com/xiewilliams/MARCO}
{https://github.com/xiewilliams/MARCO}.
\end{abstract}



\maketitle

\section{Introduction}

\begin{figure}[!t]
\centering
\includegraphics[trim=0cm 0cm 0cm 0cm, clip, width=1\linewidth]{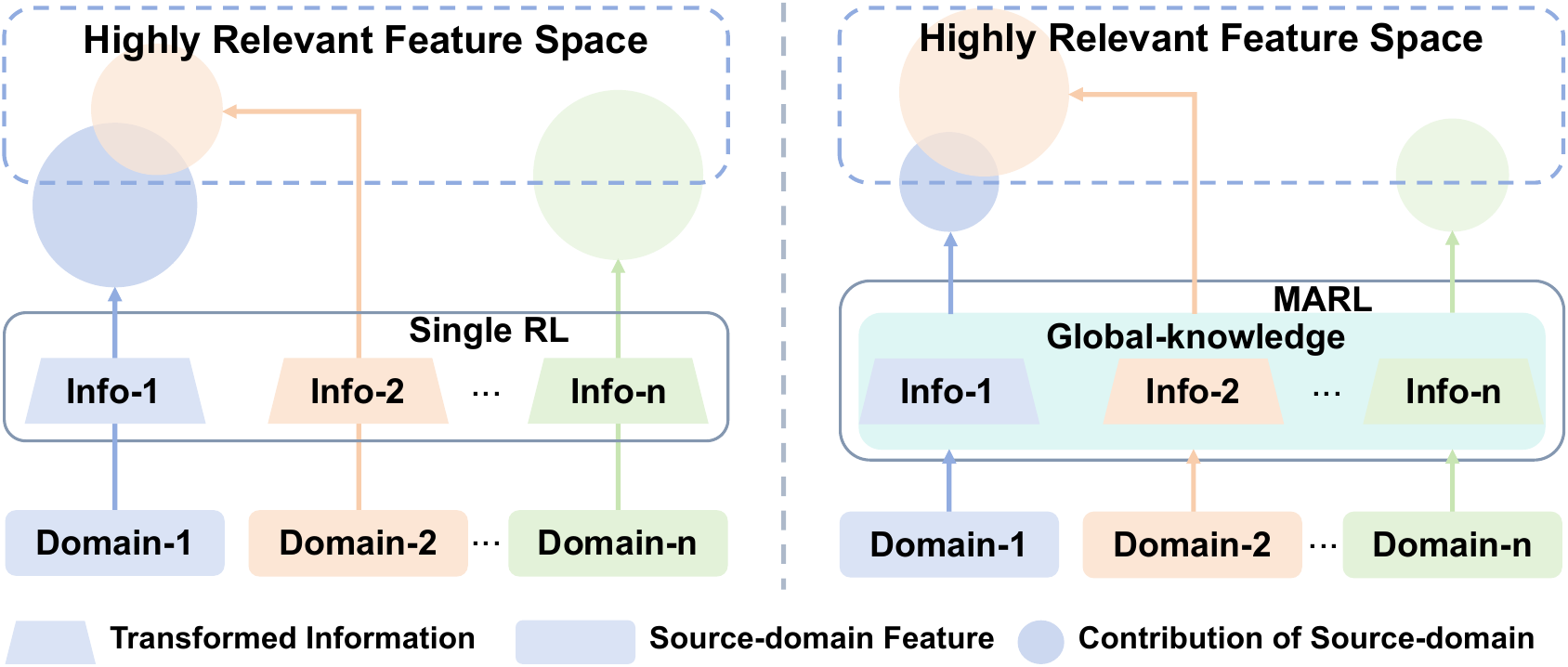}
\caption{Comparison between single-RL and MARL policies in cross-domain recommendation. Source-domain features are first transformed to personalized information. RL policies then determine each domain’s contribution to the highly relevant feature space. The radius of each circle denotes the relative contribution of each domain. In the single-RL setting, the contributions of domains 1, 2, and $n$ are inaccurately estimated due to the absence of global knowledge and cooperation optimization, whereas Multi-Agent RL yields more balanced and adaptive contributions across domains.}
\label{fig:intro}
\end{figure}

While recommender systems (RecSys) have been widely used in a variety of applications, such as e-commerce and media platforms~\cite{bpr,ncf,qiu2019rethinking,qiu2020gag,qiu2022contrastive,li2021discovering,ntk}, a significant challenge persists due to data sparsity, particularly evident when addressing cold-start issues involving new users or items~\cite{graph-cdr,llm-instructing,data-distribution-sparsity,qiu2021memory}. To mitigate this, multi-source cross-domain recommendation (CDR) has emerged as a promising paradigm, leveraging insights from multiple source domains to enhance recommendations in a target domain~\cite{emb_cdr,CATN,semi-cf,multi-mpf}. Effective CDR hinges upon accurately transferring valuable knowledge through identifying and utilizing domain dependencies at the user/item level and their interaction patterns~\cite{cdr_survey}.

Existing CDR techniques primarily focus on extracting explicit user and item features~\cite{emb_cdr,CATN,allforone,personalized—transfer-revier} and discovering interaction patterns across domains~\cite{multi-cf,cd-fm,graph-reviewer,graph-cdr}. Driven by the interactive and adaptive nature of RecSys, reinforcement learning (RL) has further advanced the CDR field by dynamically modeling user preferences~\cite{remit,reinforce_retrival,dynamic_rl_retrival}. For example, REMIT~\cite{remit} employs a \textit{single RL agent} to estimate user preferences across multiple source domains, integrating these preferences with item features via collaborative filtering to improve target-domain recommendations.

While reinforcement learning-based cross-domain recommendation methods have demonstrated promising improvements, their effectiveness remains constrained by \textbf{negative transfer}~\cite{ham,pep,allforone,negative_transfer}, primarily arising from two factors: (1) \textbf{inconsistent contributions} from multiple source domains toward the target domain and (2) \textbf{inherent discrepancies} in data distributions among source domains. Specifically, when a single agent estimates user preference associated with a particular source domain, it may overestimate the contribution of this domain if another source domain actually provides more valuable information. Conversely, the contribution of a domain may be underestimated due to similar \textbf{misassignment}. As illustrated in the left part of Figure~\ref{fig:intro}, the contribution of domain~2 is underestimated due to the overestimation of domain~1 in the single-agent setting, which fails to capture the cooperative relationships among information from different source domains, particularly without a global optimization perspective~\cite{mappo,QMIX,marl-withoutcomu,marl_recom}. Additionally, the inherent sparsity and heterogeneity of data distributions across source domains in CDR further exacerbate the difficulty of optimizing and stabilizing reinforcement learning policies for modeling user preferences.

To address the above problems, a general and effective framework, \textit{i.e.}, MARCO: \textbf{M}ulti-\textbf{A}gent \textbf{R}einforcement learning-based \textbf{C}ross-\textbf {D}omain recommendation, is proposed. 
MARCO utilizes a cooperative multi-agent reinforcement learning (MARL) paradigm, explicitly assigning each agent to estimate the contributions of individual source domains toward recommendation performance in the target domain. This multi-agent collaboration effectively manages the \textbf{credit assignment}~\cite{mappo,nguyen2018credit} process and mitigates negative transfer by leveraging global user and item features across domains. As illustrated in the right part of Figure~\ref{fig:intro}, the previously biased estimates of the contributions of domains 1, 2, and $n$ in the single-agent setting are corrected through the incorporation of global knowledge and a cooperative learning strategy. Further, an entropy-based penalty, which encourages diversity in joint agent actions, is introduced to robustly handle distributional discrepancies, thereby enhancing policy expressiveness and training stability. Extensive empirical evaluations conducted across four benchmark datasets demonstrate MARCO's superiority over state-of-the-art methods, highlighting its generalization capability and robustness. The main contributions are summarized as follows:

\begin{itemize}[leftmargin=*]
  \item The multi-source CDR task is formulated as a cooperative MARL problem, effectively managing credit assignment across source domains to significantly alleviate negative transfer.
  \item An entropy-based action diversity penalty is introduced within the MARL framework, substantially enhancing recommendation policy expressiveness and stabilizing training.
  \item Extensive experiments validate MARCO's superior performance against competitive baselines, underscoring its robustness and strong generalization ability across diverse settings.
\end{itemize}

\section{Related Work}
\subsection{Cross-domain Recommendations}
Existing CDR approaches are categorized into two categories: single-source transfer and multi-source transfer.

For the \textbf{single-source} methods, the domain with a denser interaction matrix is usually treated as the source domain and transfers the rich information to the target domain~\cite{semi-cf,cd-fm,addition-singcf-trans,MTNET,emb_cdr,pinet}. CDCF is proposed in single-source scenarios~\cite{clare,cd-fm,addition-singcf_spar} to capture shared information. Graph-based approaches learn graph representations of users and transfer them to the target domain~\cite{graph-single1,graph-single2}. The bridge-based methods are developed to transfer user preferences through mapping function across source and target domains~\cite{addition-singebridge_fusingreviews,emb_cdr,MTNET,pinet,CATN}. In contrast to these methods, MARCO addresses a more challenging multi-source CDR problem with a multi-agent RL framework.

\textbf{Multi-source} domain approaches consider combining multiple domains for target domain recommendation~\cite{multisparse,HeroGRAPH,multi-rnn,allforone,addition-mulcf_cfm,addition-mulbri-gam}. CDCF-based solutions in multiple domains~\cite{addition-mulcf_cfm,addition-mulcf_cft,addition-mulcf_tf} usually factorize and integrate latent features for the target domain. Graph-based methods construct a multi-graph based on users' behaviors from different domains and utilize multi-graph networks to learn cross-domain embedding~\cite{graph-multisource,graph-multisource2}. Bridge-based methods usually learn a shared global user embedding across domains~\cite{multi-mpf,multi-parashare,HeroGRAPH,allforone}. Different from these multi-source methods, MARCO consists of a multi-agent RL pipeline that can cooperatively learn and transfer beneficial knowledge across multiple domains.

\subsection{Reinforcement Learning in RecSys}
Generally, RL methods for recommender systems can be divided into: single-agent methods and multi-agent methods.

Most existing RL methods for recommendation are based on \textbf{single-agent RL}~\cite{pg-con,pg-text,q-drn,q-stab,DDPG-DRR,DDPG-list,DDPG-hier,yi2024roler}. Deep Q-Learning (DQN)~\cite{q-drn,q-stab} and deep deterministic policy gradient (DDPG)~\cite{DDPG-DRR,DDPG-list,DDPG-end2end,DDPG-Topaware} are the most popular single-agent frameworks. DRN~\cite{q-drn} introduces a DQN-based recommendation framework with an effective exploration strategy to address the dynamic nature of news recommendations. DRR~\cite{DDPG-DRR} makes use of deep deterministic policy gradient networks to model the long-term dynamic interactions and capture the interactions between items and users. Although previous single-agent RL methods have made some progress in recommendation systems, they lack the flexibility to address the complexities of multi-domain scenarios. Compared with these models, MARCO aims to solve the CDR problem with a multi-agent RL method in multi-domain settings.

\textbf{Multi-agent RL} can tackle tasks with higher complexity and learn more robust strategies within a shared environment than single-agent~\cite{ma_re_mul,marl-Mahrl,marl-twitter,marl-vech,marl-joint,marl-withoutcomu,yi2025darlr}. RAM introduces a two-level reinforcement learning framework to jointly optimize recommending and advertising strategies~\cite{marl-joint}. MASSA proposes a signal network to maximize the global utility and promote cooperation of all modules~\cite{marl-withoutcomu}. 
CROMA proposes a cooperative multi-agent approach to select a small set of historical tweets~\cite{marl-twitter}. MaHRL employs hierarchical multi-agent reinforcement learning to tackle multi-goal recommendations~\cite{marl-Mahrl}. Although these methods have shown advancements in handling complex tasks, they are not adaptive for multi-source cross-domain scenarios, while MARCO successfully develops a cooperative multi-agent RL framework for CDR.

\section{Preliminary}
\subsection{Problem Definition}
Following recent works~\cite{ptup,HeroGRAPH,allforone} on cross-domain recommendations, the multi-source settings are adopted. With a set of overlapping users $\mathcal{U} = \{u_1, u_2, \ldots\, u_i\}$ from various domains, a set of items $\mathcal{V} = \{v_1, v_2, \ldots\, v_j\}$, and a set of domains $\mathcal{D} = \{1, 2, \ldots\, N\}$, the user-item interactions in domain $d \in \mathcal{D}$ are represented by the rating $r_{ij} \in \mathbf{R}^{d}$. The target is to learn a personalized recommendation model to leverage knowledge from each user’s behavior sequence $\mathcal{S}_{u_i} = \{v^d_1, v^d_2, \ldots, v^d_j\}$ in all $N$ source domains and enhance the recommendation for new users and new items of the target domain, where $v^d_j$ represents the $j$-th interacted item in $d$ domain. Moreover, the items in different domains $\mathcal{V}^d$ are not shared, and interactions in a specific domain remain unobservable to other domains.

\subsection{Multi-agent Reinforcement Learning}
Multi-agent reinforcement Learning (MARL) is an extension of single-agent reinforcement learning (RL) to scenarios where multiple agents interact within the same environment. A MARL problem can be formulated as a decentralized partial observable Markov Decision Process (Dec-POMDP) with $G = \langle m, \mathcal{S}, \mathcal{A}, \mathcal{P}, \mathcal{O}, R, \gamma \rangle$ where $m$ is the number of agents and $\gamma$ is the discount factor. The global state of the environment is represented by $\mathbf{s}\in \mathcal{S}$. Each agent $i$ has a local observation $\mathbf{o}^i\in \mathcal{O}$ and takes an action $\mathbf{a}^i\in \mathcal{A}$. The joint action from all agents is denoted by \(\mathbf{A} = (\mathbf{a}^1, \ldots, \mathbf{a}^m)\) with \(\mathbf{O} = (\mathbf{o}^1, \ldots, \mathbf{o}^m)\) representing the local observations of all agents. At time $t$, each agent $i$ selects and executes an action $\mathbf{a}^i_t$ depending on the policy function $\pi_{\theta_i}(\mathbf{a}^i_t\mid\mathbf{o}^i_t)$, where $\theta_i$ is the parameter of agent $i$'s actor network. Subsequently, the environment evolves from state $\mathbf{s}_t$ under the joint action $\mathbf{A}_t$ with respect to the transition function $\mathcal{P}$ to the next global state $\mathbf{s}_{t+1}$, and each agent receives the reward from the reward function $R$.

The objective of cooperative MARL is to maximize the accumulated team reward:
\begin{equation}
J(\Theta) = \mathbb{E} \left[ \sum_{t=0}^{T} \gamma^t R(s_t, \mathbf{A}_t) \mid s_0, \mathbf{A}_t \sim \pi_{\Theta}(\cdot \mid \mathbf{O}_t) \right],
\end{equation}
where \( J(\Theta) \) is the expected accumulated reward under the joint policy parameterized by \(\Theta = \{\theta_1, \theta_2, \ldots, \theta_m\}\) and \(\pi_{\Theta}(\cdot \mid \mathbf{O}_t)\) represents the joint policy determined by the parameters \(\Theta\), conditioned on the joint observation \(\mathbf{O}_t\).

\begin{figure*}[!t]
    \centering
\begin{minipage}[b]{1\linewidth}
\centering
\includegraphics[width=\linewidth]{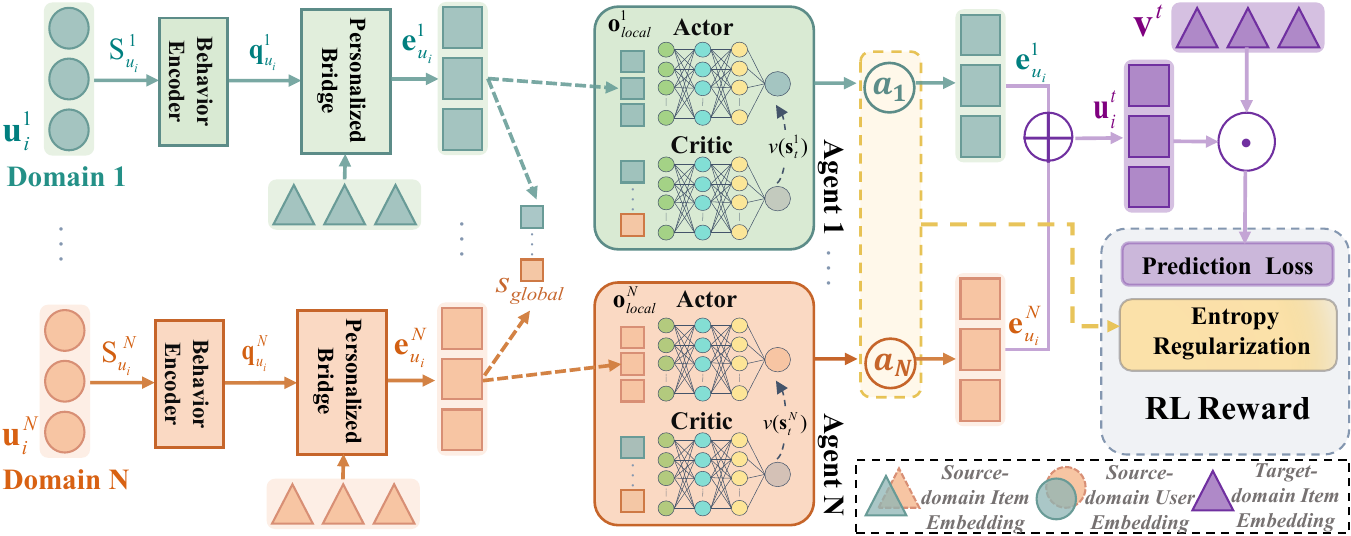}
\caption{The overview of MARCO. The domain-specific user embeddings $\mathbf{u}^d_i$ and the sequence of item embeddings $\mathcal{S}^d_{u_i}$ are trained independently through Matrix Factorization (MF). Personalized bridge modules utilize MLP networks with encoded sequence embeddings $\mathbf{q}^{d}_{u_i}$ as input to generate personalized transformed domain-specific embeddings $\mathbf{e}^{d}_{u_i}$ for user ${u}_{i}$. To leverage the most informative and transferable knowledge from source domains, the Multi-agent Proximal Policy Optimization (MAPPO) framework with a cross-domain entropy term is adopted to determine the weight of domain-specific embeddings $\mathbf{e}^{d}_{u_i}$ and obtain the initial embedding $\mathbf{u}^t_i$ for the cold start user in the target domain to boost the recommendation performance.}
\label{fig:framework}
\end{minipage}
\end{figure*}

\section{Method}
The architecture of the proposed MARCO framework for improving multi-source CDR performance is presented in Figure~\ref{fig:framework}.

\subsection{Source Domain Representation}
In each source domain, the pre-trained domain-specific embeddings are constructed through base recommendation models, which aim to capture essential features of each domain:
\begin{equation}
    \abovedisplayskip=2pt
    \belowdisplayskip=2pt
    \mathbf{v}^d_j=\text{Embed}(v_j),\quad\mathbf{u}^d_i=\text{Embed}(u_i),\label{emb}
\end{equation}
where the $j$-th item and $i$-th user are represented by $\mathbf{v}^d_j, \mathbf{u}^d_i \in \mathbb{R}^k$ denoting the $k$-dimensional representations in the $d$-th domain. $\text{Embed}(\cdot)$ can be any base model that provides embeddings, such as Matrix Factorization (MF) and NCF~\cite{ncf}. MARCO utilizes MF, one of the most widely used and foundational models in recommendation systems, as the embedding model. MF offers significant computational advantages due to its linear formulation and low parameter complexity. Furthermore, its simplicity ensures that performance differences can be clearly attributed to our proposed framework rather than the embedding model. 

\subsection{Multi-source Personalized Bridge}
The bridge-based methods usually map an overall user representation across two domains via a multi-layer perceptron~\cite{MTNET,emb_cdr,pinet}. Additionally, inspired by ~\cite{ptup, remit}, personalized bridge functions are incorporated across multiple domains to further enhance recommendation personalization.

\subsubsection{Behavior Encoder.} The behavior encoder is designed to compress the historical behaviors to a single domain representation. Since users’ historically interacted items in $\mathcal{S}_{u_i}$ contribute differently to the transferable knowledge, an attention mechanism is employed to integrate the nonlinear relationships between items and obtain a domain representation $\mathbf{q}_{u_i}^d \in \mathbb{R}^{k}$ for a specific user $u_i$. Inspired by the recent success of modeling the attention weight with neural networks~\cite{attention-net}, an MLP-based module is utilized to parameterize the attention network: $\alpha_{j}^{d'} = \text{MLP}_{att}(\mathbf{v}_j^d;\psi)$ and attentively aggregate the item embedding $\mathbf{v}_j^d \in \mathcal{S}^d_{u_i}$ in domain $d$ as follows:
\begin{equation}
    \abovedisplayskip=2pt
    \belowdisplayskip=2pt
    \mathbf{q}_{u_i}^d = \sum_{ \mathbf{v}_j^d \in \mathcal{S}_{ui}^d}\alpha_{j}^d \mathbf{v}_j^d,\quad\alpha_{j}^d 
     = \frac{\exp(\alpha_{j}^{d'})}{\sum_{ \mathbf{v}_j^d \in \mathcal{S}_{ui}^d} \exp(\alpha_{j}^{d'})},
\end{equation}
where $\psi$ denotes the parameters of $\text{MLP}_{att}(\cdot)$, and $\mathbf{q}_{u_i}^d \in \mathbb{R}^{k}$ is a transferable user embedding for user $u_i$ in the $d$-th domain. $\alpha_{j}^d$ is an attention score of item embedding $\mathbf{v}_j^d \in \mathbb{R}^{k}$.

\subsubsection{Personalized Bridge.} The common transfer function for all users struggles to capture diverse preferences across multiple source domains. Thus, personalized bridge functions are developed to transfer the distinct users' preferences across domains. A meta-network parameterized by $\eta$ is utilized to build the personalized bridge function for each domain, taking the transferable user embedding $\mathbf{q}_{u_i}^d \in \mathbb{R}^{k}$ as input. The parameters $\mathbf{W}^{d}_{u_i} \in \mathbb{R}^{k \times k}$ for the personalized bridge function of the $d$-th domain are calculated as:
\begin{equation}
    \abovedisplayskip=2pt
    \belowdisplayskip=2pt
    \mathbf{W}^{d}_{u_i} = \text{MLP}(\mathbf{q}^{d}_{u_i};\eta^{d}).
\end{equation}
Thus, the domain-specific transformed representation derived from the bridge of user $u_i$ is formulated as:
\begin{equation}
    \abovedisplayskip=2pt
    \belowdisplayskip=2pt
    \mathbf{e}^{d}_{u_i} = \text{MLP}(\mathbf{u}^{d}_{i}; \mathbf{W}^{d}_{u_i}).
\end{equation}
The transformed source information will facilitate the learning process of the Cooperative Multi-agent Estimator to generate the integration strategy for target domain recommendation.

\subsection{Cooperative Multi-Agent Estimator}
The cooperative multi-agent estimator is designed to incorporate domain-specific characteristics. To ensure generalization capacity, the cooperative MARL algorithm is implemented as the multi-agent proximal policy optimization (MAPPO)~\cite{mappo}, which employs actor-critic networks and a clipping mechanism to stabilize policy optimization. Specifically, each agent involves its own actor and critic networks, where the critic evaluates the actions chosen by the actor and the actor outputs actions aimed at maximizing the global cumulative reward. By updating the actor network based on the feedback of the critic and updating the critic network based on the feedback of the environment, each agent learns to dynamically generate personalized actions that serve as the weight to integrate transformed source embeddings. The key items of the Dec-POMDP scenario are illustrated as follows:

\textbf{Observation} \(\mathbf{O} = (\mathbf{o}_{local}^1, \ldots, \mathbf{o}_{local}^m)\) denotes the set of local observations used by agents to choose actions. For the agent $i$, the local observation $\mathbf{o}_{local}^{i} \in \mathbb{R}^{1 \times (2k+1)}$ which contains the transformed domain embedding $\mathbf{e}^{d}_{u_i} \in \mathbb{R}^{k}$ from the $d$-th personalized bridge function, pre-trained item embedding $\mathbf{v}^{t} \in \mathbb{R}^{k}$ in the target domain and the prediction vector $\mathbf{h}_{local}^{t} \in \mathbb{R}^{1 \times 1}$ that is derived from a dot product between $\mathbf{e}^{d}_{u_i} \in \mathbb{R}^{k}$ and $\mathbf{v}^{t} \in \mathbb{R}^{k}$:
\begin{equation}
    \abovedisplayskip=2pt
    \belowdisplayskip=2pt
    \label{local_state}
    \mathbf{o}_{local}^{i} = [\mathbf{e}^{d}_{u_i}\|\mathbf{v}^{t}\|\mathbf{h}_{local}^{t}],\quad\mathbf{h}_{local}^{t} = \mathbf{e}^{d}_{u_i} \cdot \mathbf{v}^{t},
\end{equation}
where $\|$ represents a concatenation operation.

\textbf{Global State} $\mathbf{s}$ is formed by concatenating all local agent observations~\cite{mappo}. To obtain the global state $\mathbf{s}_{global}\in \mathbb{R}^{k}$, target-aware attention is adopted to selectively focus on the most relevant information of each domain:
\begin{equation}
    \abovedisplayskip=2pt
    \belowdisplayskip=2pt
    \mathbf{s}_{global} = \sum\nolimits_{d=1}^{N} \alpha^{d}\mathbf{e}^{d}_{u_i},
    \label{global_state}
\end{equation}
where $\alpha^{d}$ is the attention score of the correlations between the target domain and source domain embeddings.

\textbf{Action} \(\mathbf{A} = (\mathbf{a}^1, \ldots, \mathbf{a}^m)\) is a set of actions for each agent, where $\mathbf{a}^i$ is the action of the agent $i$. The action of each agent is defined as the weight in Eq.~(\ref{target_emb}) of each transformed domain-specific embedding and learned by the actor network, which takes the agent’s local observation as input.

\textbf{Reward} $R$ is the global reward function defined as the negative of the ground truth loss. When agents take action, the integrated transformed embedding in Eq.~(\ref{target_emb}) of the target domain will be fed into the recommender to train the bridge function module based on Eq.~(\ref{reward}). Lower ground truth loss corresponds to a higher reward, the learning process aims to achieve more accurate predictions.

\subsubsection{Actor network.}
The actor network parameterized by $\theta$ generates the action for each agent to collaboratively maximize the accumulated team reward. At each timestep $t$, the actor utilizes the local observation as input to generate the action for each agent $i$ with knowledge obtained from source domains as follows: 
\begin{equation}
    \abovedisplayskip=2pt
    \belowdisplayskip=2pt
    a_t^i = \text{Actor}(\mathbf{o}^i_{local}; \theta_i).
\label{actor}
\end{equation}
The actions will be implemented as weights of domain-specific transformed embedding, where $0 \leq a_t^i \leq 1$ and $\sum_{i=1}^{m} a_t^i = 1$ (m is the number of agents). Subsequently, the weighted sum is used to calculate the initial embedding of new users in the target domain in Eq.~(\ref{target_emb}) for prediction.

\subsubsection{Critic Network.} 
The critic network parameterized by $\phi$ evaluates the state value ${v}(\mathbf{s}_t)$ for the action generated by the actor to help update the actor network in the direction of producing better actions. At each timestep $t$, the local state and the global state embeddings are concatenated: $\mathbf{s}_t^i = [\mathbf{o}_{local}^i \| \mathbf{s}_{global}]$ as input to estimate the state value $v(\mathbf{s}_t^i)$ for agent $i$ as follows:
\begin{equation}
    \abovedisplayskip=2pt
    \belowdisplayskip=2pt
     {v}(\mathbf{s}_t^i) = \text{Critic}(\mathbf{s}_t^i\ ; \phi_i).
\label{critic}
\end{equation}
In addition, the structures of the critic network and the actor network will be described in the subsequent experimental section.

\subsection{Entropy-based Action Diversity Penalty} 
Although MAPPO inherently incorporates an entropy penalty, it would be proved not as effective as the entropy penalty introduced in this part in the experiment section. In the forward process, action values between 0 and 1 are determined by actor networks, which aim to generate \(a_t^i\) with \(\sum_{i=1}^{m} a_t^i = 1\) as the weight of transformed embeddings. The set of actions can be interpreted as a probability distribution, and the distinction of actions can be measured with entropy. A higher entropy value indicates that the distribution is more uniform, meaning each domain-specific transformed embedding has a similar significance. In contrast, a lower entropy value suggests that the distribution is more distinct, with some domains having more essential information than others. Based on these insights, the cross-domain entropy-based action diversity penalty is formulated as:
\begin{equation}
    \abovedisplayskip=2pt
    \belowdisplayskip=2pt
    H_{\text{joint}} = -\sum\nolimits_{i=1}^{m} a^i_t \log(a^i_t),
\label{entropy_action}
\end{equation}
where $H_{\text{joint}}$ is the actions entropy and $a^i_t$ is the chosen action of agent $i$ at time $t$. Therefore, the objective function derived from MAPPO~\cite{mappo} for updating the actor network of agent $i$ is:
\begin{equation}
    \abovedisplayskip=2pt
    \belowdisplayskip=2pt
    \label{entropy_loss}
    \begin{aligned}
    \ell(\theta_i) &= \frac{1}{m} \sum_{i=1}^{m}\hat{\mathbb{E}}_t \bigg[ \min(r^i_t(\theta_i) \hat{A}^i_t, \text{clip}(r^i_t(\theta_i), 1-\epsilon, 1+\epsilon) \hat{A}^i_t)\\
    &- \beta H_{\text{joint}} \bigg],\quad\text{where } r_t^i(\theta_i) = \frac{\pi_{\theta_i}(a_t^i | \mathbf{o}^i_{local})}{\pi_{\theta_{i, \text{old}}}(a_t^i | \mathbf{o}^i_{local})},
    \end{aligned}
\end{equation}
where $\beta$ is the coefficient that balances the importance of entropy regularization in the objective function. \textbf{With the entropy term $H_{\text{joint}}$, the actor network can be trained to prefer action distribution with lower entropy, leveraging the most informative and transferable embeddings from source domains.} $r_t^i(\theta_i)$ is defined as the importance ratio of the probability for taking the action \( a_t^i \) under the current policy to the old policy. $\hat{\mathbb{E}}_t$ denotes the empirical expectation over a finite batch of samples. Following MAPPO~\cite{mappo}, the clip rate $\epsilon$ is utilized to stabilize the learning process by constraining the importance ratio $r_t^i(\theta_i)$ within the range $[1 - \epsilon, 1 + \epsilon]$, and $\hat{A}_t^i$ is the estimator of the advantage function to reduce the variance of the advantage estimation within a trajectory~\cite{PPO}.

The clipped objective of MAPPO~\cite{mappo} is also utilized to update the critic network for the agent 
$i$ to achieve a trade-off between training stability and learning efficiency:
\begin{equation}
    \abovedisplayskip=2pt
    \belowdisplayskip=2pt
    \begin{split}
    &\ell_\text{Critic} = \frac{1}{m} \sum\nolimits_{i=1}^{m} \hat{\mathbb{E}}_t \Bigg[ \max \Bigg( \left( v_{\phi_i}(\mathbf{s}^i_t) - v^i_{\text{targ}}(\mathbf{s}^i_t) \right)^2, \\
    &\left( \text{clip}\left( v_{\phi_i}(\mathbf{s}^i_t), v_{\phi_{i, \text{old}}}(\mathbf{s}^i_t) - \epsilon, \right. \right.\left. \left. v_{\phi_{i,\text{old}}}(\mathbf{s}^i_t) + \epsilon \right) - v^i_{\text{targ}}(\mathbf{s}^i_t) \right)^2 \Bigg) \Bigg],
    \end{split}
    \label{critic_loss}
\end{equation}
where $v_{\phi_i}(\mathbf{s}^i_t)$ represents the predicted state value of agent $i$ with parameters $\phi_i$ at time $t$, and $v_{\text{targ}}(\mathbf{s}^i_t)$ is the target value based on the reward received from the environment.

To reduce computational complexity and improve scalability, parameter sharing is applied across agents for both the actor and critic networks~\cite{mappo}, enabling efficient learning while maintaining coordination among agents.
\begin{algorithm}[!t]
\caption{MARCO for Cross-domain Recommendations}
\label{alg:Framwork}
\begin{algorithmic}[1] 
\REQUIRE ~~Training dataset $\mathcal{D}$; Pre-trained basic recommender; Total training epoch $K$; Multi-source personalized bridge network (MPB) parameterized by $\eta$; Actor network parameterized by $\theta$; Critic network parameterized by $\phi$; Replay buffer $\mathcal{R}$
    \ENSURE ~~Learned policy $\pi_\theta$ for recommendations
   
    \STATE Initialize parameters ($\eta$,$\theta$,$\phi$) and the replay buffer $\mathcal{R}$;
    \STATE Train the basic recommender by Eq.~\eqref{emb};
    \FOR{epoch $k = 0,1,2, ..., K-1$}
        \STATE \# Update replay buffer $\mathcal{R}$ and MPB
        \FOR{batch in $\mathcal{D}$}
            \STATE Obtain $\mathbf{s_t}$ with $\mathbf{o}^i_{local}$, $\mathbf{s}_{global}$ by MPB via Eq.~\eqref{local_state} and~\eqref{global_state}.
            \STATE Sample an action for each agent $a^i_t = \pi_i(\mathbf{o}^i_{local};\theta_t^i)$ as the weight of each domain.
            \STATE Calculate the transformed embedding $\mathbf{u}^t_i$ in target domain by Eq.~\eqref{target_emb}.
            \STATE Calculate the reward $\mathbf{r_t}$ by Eq.~\eqref{reward}.
            \STATE Store the trajectory $\tau = \{(\mathbf{s_0}, a_0, r_1), (\mathbf{s_1}, a_1, r_2), \ldots)\}$ in replay buffer $\mathcal{R}$.
            \STATE Update the parameter $\eta$ of the MPB model.
        \ENDFOR
        \STATE \# Update Actor and Critic networks
        \FOR{each trajectory in replay buffer $\mathcal{R}$}
            \STATE Calculate action for each agent by Eq.~\eqref{actor}.
            \STATE Calculate state value by Eq.~\eqref{critic}.
            \STATE Calculate the entropy of all agents by Eq.~\eqref{entropy_action}.
            \STATE Update the actor, \textit{i.e.,} $\theta$, by Eq.~\eqref{entropy_loss}.
            \STATE Update the critic, \textit{i.e.,} $\phi$, by Eq.~\eqref{critic_loss}.
             
        \ENDFOR
    \ENDFOR
\RETURN Recommendation policy $\pi_\theta$;
\end{algorithmic}
\end{algorithm}

\subsection{Model Inference}
The algorithm is shown in Algorithm 1. Firstly, the MF model is applied to obtain domain-specific embeddings: $\mathbf{u}^d_i\in \mathbb{R}^{k}$ and $\mathbf{v}^d_j\in \mathbb{R}^{k}$. Subsequently, the multi-source personalized bridge module takes the pre-trained embeddings as input to learn transformed user preference $\mathbf{e}^{d}_{u_i}\in \mathbb{R}^{k}$ for different source domains. Finally, the multi-agent reinforcement estimator framework is used to integrate transformed domain-specific embeddings as follows:
\begin{equation}
    \abovedisplayskip=2pt
    \belowdisplayskip=2pt
    \mathbf{u}^t_i = \sum\nolimits_{d=1}^{N} {p}^{d}_{u_i} \cdot \mathbf{e}^{d}_{u_i},
    \label{target_emb}
\end{equation}

where ${p}^{d}_{u_i}$ represents the value derived from the multi-agent actor network's output. The personalized initial embedding for the user $u_i$ who is a new user in the target domain is denoted as $\mathbf{u}^t_i\in\mathbb{R}^{k}$. Based on the comprehensive initial embedding in the target domain, the training procedure focuses on optimizing the rating prediction task for target recommendation as follows:
\begin{equation}
    \abovedisplayskip=2pt
    \belowdisplayskip=2pt
    \min\frac{1}{|\mathcal{W}^{t}|} \sum_{w_{u_i}\in \mathcal{W}^{t}} (\mathbf{u}^t_i\cdot \mathbf{v}^{t} - {w}_{u_i})^2,
    \label{reward}
\end{equation}
where $\mathcal{W}^{t}$ is the ground truth for the rating task in the target domain and $|\mathcal{W}^{t}|$ is the size of the set $\mathcal{W}^{t}$.

\section{Experiment}
In this section, extensive experiments are conducted to answer the following research questions (RQs):
\begin{itemize}[leftmargin=*]
    \item \textbf{RQ1}: How does MARCO perform compared with state-of-the-art CDR methods?
    \item \textbf{RQ2}: How do the components and designs of MARCO contribute to overall performance?
    \item \textbf{RQ3}: How does the generalization capacity of MARCO vary across various cold start rates? 
    \item \textbf{RQ4}: How does the robustness of MARCO vary across various numbers of source domains?
    \item \textbf{RQ5}: What is the sensitivity of hyperparameter $\beta$ in MARCO?
    \vspace{-2em}
\end{itemize}

\textcolor{red}{\begin{table}[!t]
    \caption{Dataset statistics.}
    \centering
    \begin{tabular}{ccccc}
    \toprule
    Domain & \#Users & \#Overlap Users & \#Items & \#Interactions \\
    \midrule
    CD   & 75{,}258   & 2{,}166   & 64{,}443   & 1{,}097{,}592  \\
    MV   & 123{,}960  & 2{,}166   & 50{,}052   & 1{,}697{,}533  \\
    Book & 603{,}668  & 2{,}166   & 367{,}982  & 8{,}898{,}041  \\
    Elec & 725{,}523  & 2{,}166   & 63{,}001   & 1{,}689{,}188  \\
    \bottomrule
    \end{tabular}
    \label{table:dataset_info}
    \vspace{-0.5em}
    \end{table}
}

\textcolor{red}{
\begin{table}[!t]
    \caption{Computational complexity.}
    \vspace{-0.5em}
    \centering
    \begin{tabular}{lcc}
    \toprule
    Model & MARCO & REMIT \\
    \midrule
    Total parameters (M) & 106.86 & 73.06 \\
    FLOPs/sample (G) & 2.06 & 1.44 \\
    Training time (GPU-h) & 14 & 13 \\
    Inference latency/sample (ms) & 0.06 & 0.04 \\
    \bottomrule
    \end{tabular}
    \label{model_efficiency}
    \vspace{-0.5em}
\end{table}
}

\subsection{Setup}
\subsubsection{Datasets}
Following~\cite{ptup,remit,emb_cdr}, a real-world public dataset named Amazon review dataset\footnote{https://jmcauley.ucsd.edu/data/amazon/}is used for evaluation. Each user or item has at least five ratings, and the statistics of the dataset are shown in Table~\ref{table:dataset_info}. Multi-source recommendation tasks are conducted on the four sub-category datasets: cds\_and\_vinyl (CD), electronics (Elec), movies\_and\_tv (MV), and books (Book). Domains are linked through common user IDs. Each category takes turns as the target domain, with the others serving as source domains.

\subsubsection{Metrics}
Root Mean Square Error (RMSE) and Mean Absolute Error (MAE) are employed to evaluate the predictive accuracy of recommendations given the explicit rating scores (0-5).

\subsubsection{Baselines}
The following cross-domain methods serve as baselines of the proposed MARCO:
\begin{itemize}
    [leftmargin=*]
    \item 
    \textbf{MF}~\cite{bpr} is a recommendation method trained only with information from the target domain.
    \item \textbf{CMF}~\cite{CMF} (\textbf{\textit{MF-based}}) is an extension of MF for multiple domains and jointly factorizes multiple related matrices by sharing latent representations for common entities.
    \item \textbf{EMCDR}~\cite{emb_cdr} (\textbf{\textit{Bridge-based}}) learns domain-specific features and utilizes nonlinear mapping functions to transfer user information.
    \item \textbf{PTUP}~\cite{ptup} (\textbf{\textit{Bridge-based}}) learns users' characteristic embeddings and builds personal mapping functions to bridge user embeddings from the source to the target domain.
    \item \textbf{REMIT}~\cite{remit} (\textbf{\textit{Single-agent RL}}) is a CDR method using REINFORCE~\cite{reinforce} to learn multi-interests and transferred embeddings.
\end{itemize}
\begin{table}[!t]
  \centering
  \caption{Overall results (MAE and RMSE) for three cold start scenarios. The best results are in boldface. Imp. (\%) denotes the relative improvement over the best baseline. The results are reproduced according to the authors' open-source code.}
    \resizebox{1\linewidth}{!}{
    \begin{tabular}{c|c|c|ccccc|c|c}
    \toprule
         & Target & Metric & MF & CMF & EMCDR & PTUP & REMIT & MARCO (ours) & Imp.\\
    \midrule 
    \multirow{8}[8]{*}{\rotatebox[origin=c]{90}{20\% Cold Start Rate}} & \multirow{2}[2]{*}{CD} & MAE & 4.363 & 1.840 & 1.263 & 1.116 & \underline{1.056} & \textbf{0.823 $\pm$ 0.007} & 22.0\\
          &       & RMSE & 5.105 & 2.357 & 1.617 & 1.479 & \underline{1.397} & \textbf{1.087 $\pm$ 0.004} & 22.1 \\
          \cmidrule{2-10} 
          & \multirow{2}[2]{*}{MV} & MAE & 4.162 & 1.404 & 1.037 & 1.006 & \underline{0.959} & \textbf{0.858 $\pm$ 0.004} & 10.5 \\
          &       & RMSE & 4.741 & 1.683 & 1.381 & 1.298 & \underline{1.244} & \textbf{1.109 $\pm$ 0.002} & 10.8 \\
          \cmidrule{2-10} 
          & \multirow{2}[2]{*}{Book} & MAE & 4.512 & 1.785 & 1.155 & 1.136 & \underline{1.111} & \textbf{0.925 $\pm$ 0.005} & 16.7 \\
          &       & RMSE & 5.137 & 2.349 & 1.452 & 1.446 & \underline{1.426} & \textbf{1.171 $\pm$ 0.002} & 17.9 \\
          \cmidrule{2-10} 
          & \multirow{2}[2]{*}{Elec} & MAE & 4.300 & 2.238 & 1.242 & 1.150 & \underline{1.052} & \textbf{0.980 $\pm$ 0.004} & 6.8 \\
          &       & RMSE & 4.823 & 2.651 & 1.555 & 1.493 & \underline{1.339} & \textbf{1.258 $\pm$ 0.003} & 6.0 \\
\midrule
    \multirow{8}[8]{*}{\rotatebox[origin=c]{90}{50\% Cold Start Rate}} & \multirow{2}[2]{*}{CD} & MAE & 4.473 & 2.280 & 1.397 & 1.151 & \underline{1.141} & \textbf{0.769 $\pm$ 0.011 } & 32.6 \\
          &       & RMSE & 5.296 & 2.828 & 1.726 & 1.524 & \underline{1.430} & \textbf{1.022 $\pm$ 0.005} & 28.5 \\
          \cmidrule{2-10}
          & \multirow{2}[2]{*}{MV} & MAE & 4.097 & 1.300 & 1.050 & \underline{1.038} & 1.065 & \textbf{0.884 $\pm$ 0.008 } & 14.8 \\
          &       & RMSE & 4.662 & 1.687 & 1.332 & \underline{1.332} & 1.348 & \textbf{1.125 $\pm$ 0.006} & 15.5 \\
          \cmidrule{2-10}
          & \multirow{2}[2]{*}{Book} & MAE & 4.299 & 1.932 & 1.178 & \underline{1.123} & 1.394 & \textbf{0.952 $\pm$ 0.001} & 15.2 \\
          &       & RMSE & 4.956 & 2.483 & 1.474 & \underline{1.433} & 1.690 & \textbf{1.212 $\pm$ 0.001} & 15.4 \\
          \cmidrule{2-10}
          & \multirow{2}[2]{*}{Elec} & MAE & 4.326 & 2.191 & 1.255 & \underline{1.158} & 1.270 & \textbf{0.941 $\pm$ 0.011} & 18.8 \\
          &       & RMSE & 4.850 & 2.625 & 1.570 & \underline{1.502} & 1.571 & \textbf{1.203 $\pm$ 0.006} & 19.9 \\
\midrule
    \multirow{8}[8]{*}{\rotatebox[origin=c]{90}{80\% Cold Start Rate}} & \multirow{2}[2]{*}{CD} & MAE & 4.504 & 2.151 & 1.492 & \underline{1.210} & 1.223 & \textbf{0.850 $\pm$ 0.008 } & 29.8 \\
          &       & RMSE & 5.169 & 2.711 & 1.835 & 1.605 & \underline{1.511} & \textbf{1.126 $\pm$ 0.012 } & 29.9 \\
          \cmidrule{2-10}
          & \multirow{2}[2]{*}{MV} & MAE & 4.180 & 1.434 & 1.266 & \underline{1.058} & 1.326 & \textbf{0.907 $\pm$ 0.010} & 14.2 \\
          &       & RMSE & 4.793 & 1.836 & 1.572 & \underline{1.356} & 1.615 & \textbf{1.174 $\pm$ 0.005 } & 13.4 \\
          \cmidrule{2-10}
          & \multirow{2}[2]{*}{Book} & MAE & 4.311 & 1.792 & 1.363 & \underline{1.145} & 1.465 & \textbf{0.952 $\pm$ 0.003} & 16.8\\
          &       & RMSE & 4.957 & 2.337 & 1.695 & \underline{1.457} & 1.828 & \textbf{1.224 $\pm$ 0.009 } & 15.9 \\
          \cmidrule{2-10}
          & \multirow{2}[2]{*}{Elec} & MAE & 4.300 & 2.114 & 1.331 & \underline{1.194} & 1.486 & \textbf{1.027 $\pm$ 0.003} & 14.0 \\
          &       & RMSE & 4.820 & 2.557 & 1.668 & \underline{1.549} & 1.804 & \textbf{1.324 $\pm$ 0.008} & 14.5 \\
    \bottomrule
    \end{tabular}%
    }
  \label{overall performance}%
\end{table}%

\subsubsection{Implementations}
All experiments are conducted on two NVIDIA GeForce RTX\texttrademark\ 4090 with 24 GB GDDR6X. For pre-training, the MF is utilized with the embedding size chosen from $\{5,10,20\}$. For the knowledge transfer module, the initial learning rate for the Adam optimizer~\cite{adam} is chosen from $\{0.001,0.01,0.1\}$ and the weight decay is from $\{0,0.0001\}$. The latent dimension of MLP is chosen from $\{20,50,80,128\}$. For the MAPPO module, the actor and critic networks are implemented as separate neural network architectures, each containing three fully connected layers followed by Tanh activation functions. The actor network generates actions through a Softmax output layer, and the critic network generates the estimated state value through a linear output layer. The discount factor $\lambda$ is chosen from $\{0.95,0.98,0.99\}$ and the clipping rate $\epsilon$ of the surrogate objective is chosen from $\{0.15,0.2,0.5\}$. The coefficient $\beta$ of entropy terms is chosen from $\{0.0001,0.001,0.15,1.5\}$ and the clipping rate of the gradient is chosen from $\{0.2,0.5\}$. The mean and the standard deviation of five random runs are reported.

A comparison of computational complexity, as shown in Table~\ref{model_efficiency}, is made between MARCO and REMIT~\cite{remit}, a single-agent RL baseline designed for cross-domain recommendation, which is adapted to the multi-source setting in our implementation. Although the MARCO parameter is moderately larger than REMIT, MARCO's training time remains comparable to REMIT (14 GPU hours vs 13 GPU hours), and its online inference latency is only marginally higher (0.06 ms vs 0.04 ms per sample). Thanks to effective parameter sharing and parallelization technologies, MARCO incurs only a modest increase in computational complexity compared to REMIT, yet delivers substantial performance gains.

Following~\cite{emb_cdr,ptup,remit}, to evaluate the effectiveness of MARCO on cross-domain recommendation, a random selection is adopted to build test data. Specifically, a fraction of overlapping users in the target domain is randomly removed to serve as test users, while the remaining overlapping user data is used for training. In this study, the proportions of test (cold-start) users are set to 20\%, 50\%, and 80\% of the total overlapping users. The model evaluation process is divided into two steps: (1) training the model on the training set and (2) testing its performance on the cold-start test set.
\vspace{-0.8em}

\subsection{Overall Performance (RQ1)}
The overall performance of all baselines and MARCO is evaluated on four Amazon datasets in cold start scenarios. The results in Table~\ref{overall performance} demonstrate the performance on four target domains under different cold rate settings. The best performance is shown in boldface, and improvement denotes the relative improvement over the best baseline. The following findings from the experimental results are: (1) As the cold rate increases, the performance of all models tends to decline due to less training data. MF performs the worst as it relies solely on target domain data. (2) Bridge-based methods like EMCDR and PTUP outperform CMF by transferring efficient knowledge through bridge-mapping functions. For instance, PTUP outperforms CMF by 34\% and MF by 71\% at a 20\% cold rate in RMSE, and by 42\% and 74\% at a 50\% cold rate in MAE, demonstrating that transferring individual user preferences achieves superior performance. (3) MARCO significantly outperforms all baselines by a large margin due to its MARL approach.

\textbf{MARCO consistently achieves superior performance to the single-RL method}. The key lies in MARCO's ability to share information and parameters across diverse domains, with each domain handled by one agent and dynamically adopting cooperative knowledge transfer strategies for personalized users, even with limited data. Such cooperation allows MARCO to fully exploit user and item features from multiple source domains. Additionally, by incorporating entropy regularization into the multi-agent framework, distinct contributions from different domains are encouraged to boost recommendation performance in the target domains.

\subsection{Ablation Study (RQ2)}
The MARCO framework has the following key components: the multi-source personalized bridge (MPB), the MARL framework, and the entropy regularization. To validate the effectiveness of these modules, the corresponding modules are removed for the ablation study, and the results are in Table~\ref{ablation table}. Specifically, the following variants are compared with MARCO: (1) \textbf{w/o MPB}: all source domains share a common personalized bridge function and maintain the MARL framework with the entropy term. (2) \textbf{REMIT}: a single-agent RL algorithm, REINFORCE~\cite{reinforce}, with a policy-gradient optimization is used. (3) \textbf{w/o MARL}: the MAPPO is replaced by the single-agent PPO~\cite{PPO}, which utilizes actor-critic networks and a clipping mechanism in single-agent optimization. (4) \textbf{w/o Ent}: using MAPPO~\cite{mappo} without the entropy regularization.

As shown in Table~\ref{ablation table}, removing any components from MARCO results in a significant decline in both RMSE and MAE, and the complete MARCO outperforms all other variants. Specifically, a single common bridge struggles to capture diverse user preferences across diverse domains. REMIT's performance drops sharply with increasing cold rates, highlighting its sensitivity to limited training data. For w/o MARL, the clipping mechanism and actor-critic networks help stabilize the PPO's training process by preventing large updates to the policy, leading to better performance but still lower than multi-agent models. The cooperative nature and shared learning in MAPPO allow for more robust and efficient policy updating than single-agent methods. The complete MARCO makes significant improvements across all cold start scenarios, proving that cross-domain entropy regularization can enhance the transfer of diverse knowledge from multiple domains.

\textbf{To evaluate the importance of entropy regularization}, the following variants are investigated: (1) \textbf{PPO}: single-agent PPO without entropy regularization. (2) \textbf{PPO w/Ent}: single-agent PPO with entropy regularization. (3) \textbf{MAPPO}: MAPPO without the entropy regularization. (4) \textbf{MAPPO w/Ent}: MAPPO with the entropy regularization. The results in Table~\ref{entropy table} indicate that entropy regularization can improve the performance in all scenarios and effectively reduce negative transfer. For the single-agent PPO, the model with entropy achieves an average 7\% improvement in MAE, while the multi-agent MAPPO achieves an average 9\% improvement, showing that entropy regularization with the multi-agent framework offers a more substantial improvement in tackling complicated tasks than the single agent. Notably, while MAPPO with entropy including a bonus for random exploration achieves good results, MARCO achieves the best performance.

\begin{figure}[!t]
    \centering
    \includegraphics[width=0.97\linewidth]{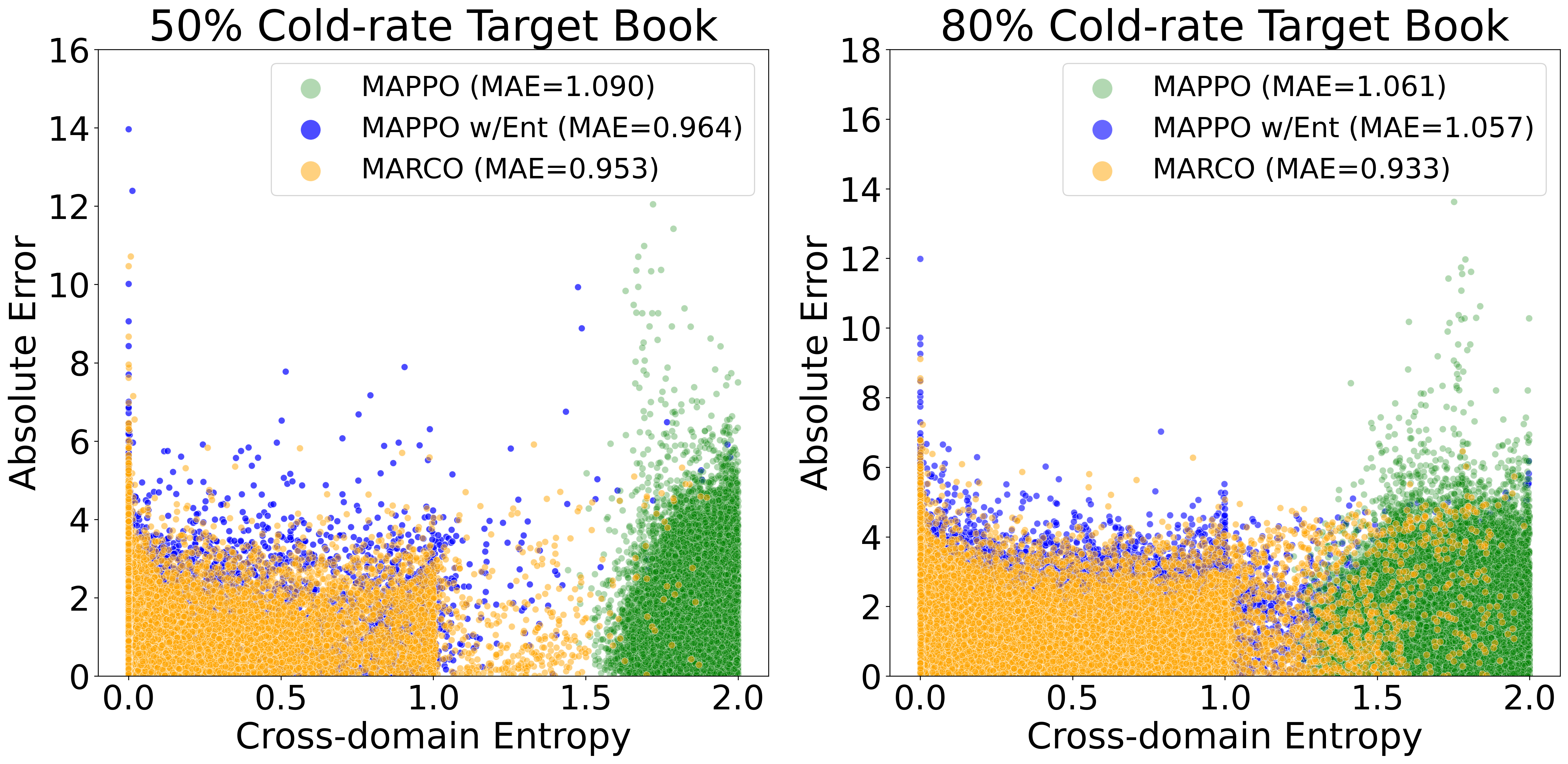}
    \caption{Visualization of the effectiveness of entropy regularization for the Book domain as the target by three variants: MARCO (\textcolor{orange}{orange}), MAPPO (\textcolor{green}{green}), and MAPPO w/Ent (\textcolor{blue}{blue}). The entropy regularization in MAPPO leads to a lower prediction error for \textcolor{orange}{orange dots} in the bottom left corner.}
    \label{entropy-vis}
    \vspace{-1em}
\end{figure}

\textbf{To visualize the effectiveness of entropy regularization}, MARCO, MAPPO w/Ent, and original MAPPO are compared to visualize the entropy for each user, as shown in Figure~\ref{entropy-vis}. With the entropy regularization, the \textcolor{orange}{orange dots} for MARCO achieve the lowest MAE error while maintaining a low entropy of the RL actions. When adding too much exploration as shown in the \textcolor{blue}{blue dots} in MAPPO w/Ent, the model will lead to a higher inaccuracy in the prediction. This is because its exploration bonus encourages agents to continuously explore diverse actions with uncertainty, resulting in less focus on important information and unstable performance. Comparing the original MAPPO (\textcolor{green}{green}) with the w/Ent (\textcolor{blue}{blue}) variant, it is clear that the green dots without the entropy regularization lead to a higher MAE error. MAPPO w/Ent alleviates the negative transfer and instability of MAPPO by incorporating an entropy regularization term, but it still retains its exploration bonus, leading to worse performance than MARCO.

\begin{table}[!t]
  \centering
  \caption{Ablation study.}
  \vspace{-0.5em}
    \resizebox{1\linewidth}{!}{
    \begin{tabular}{c|c|c|ccccc}
    \toprule
         & Target & Metric & w/o MPB & REMIT & w/o MARL & w/o Ent & MARCO\\
\midrule 
    \multirow{8}[8]{*}{\rotatebox[origin=c]{90}{20\% Cold Start Rate}} & \multirow{2}[2]{*}{CD} & MAE & 0.923 $\pm$ 0.010 & 1.056 $\pm$ 0.036 & 1.088 $\pm$ 0.013 & 0.936 $\pm$ 0.017 & \textbf{0.823 $\pm$ 0.007} \\
          &       & RMSE & 1.242 $\pm$ 0.004 & 1.397 $\pm$ 0.035 & 1.361 $\pm$ 0.022 & 1.195 $\pm$ 0.006 & \textbf{1.087 $\pm$ 0.004} \\
          \cmidrule{2-8} 
          & \multirow{2}[2]{*}{MV} & MAE & 0.924 $\pm$ 0.011 & 0.959 $\pm$ 0.027 & 0.939 $\pm$ 0.009 & 0.916 $\pm$ 0.004 & \textbf{0.858 $\pm$ 0.004} \\
          &       & RMSE & 1.187 $\pm$ 0.005 & 1.244 $\pm$ 0.037 & 1.216 $\pm$ 0.006 & 1.188 $\pm$ 0.003 & \textbf{1.109 $\pm$ 0.002} \\
          \cmidrule{2-8} 
          & \multirow{2}[2]{*}{Book} & MAE & 0.961 $\pm$ 0.006 & 1.111 $\pm$ 0.011 & 1.118 $\pm$ 0.006 & 1.056 $\pm$ 0.007 & \textbf{0.925 $\pm$ 0.005} \\
          &       & RMSE & 1.234 $\pm$ 0.019 & 1.426 $\pm$ 0.011 & 1.403 $\pm$ 0.009 & 1.354 $\pm$ 0.002 & \textbf{1.171 $\pm$ 0.002} \\
          \cmidrule{2-8} 
          & \multirow{2}[2]{*}{Elec} & MAE & 1.033 $\pm$ 0.007 & 1.052 $\pm$ 0.014 & 1.055 $\pm$ 0.008 & 1.001 $\pm$ 0.005 & \textbf{0.980 $\pm$ 0.004} \\
          &       & RMSE & 1.322 $\pm$ 0.003 & 1.339 $\pm$ 0.016 & 1.354 $\pm$ 0.008 & 1.296 $\pm$ 0.008 & \textbf{1.258 $\pm$ 0.003} \\
\midrule
    \multirow{8}[8]{*}{\rotatebox[origin=c]{90}{50\% Cold Start Rate}} & \multirow{2}[2]{*}{CD} & MAE & 1.052 $\pm$ 0.039 & 1.141 $\pm$ 0.051 & 1.127 $\pm$ 0.012 & 0.924 $\pm$ 0.009 & \textbf{0.769 $\pm$ 0.011} \\
          &       & RMSE & 1.407 $\pm$ 0.031 & 1.430 $\pm$ 0.027 & 1.375 $\pm$ 0.011 & 1.186 $\pm$ 0.011 & \textbf{1.022 $\pm$ 0.005} \\
          \cmidrule{2-8}
          & \multirow{2}[2]{*}{MV} & MAE & 0.943 $\pm$ 0.029 & 1.065 $\pm$ 0.061 & 1.006 $\pm$ 0.005 & 0.900 $\pm$ 0.006 & \textbf{0.884 $\pm$ 0.008} \\
          &       & RMSE & 1.220 $\pm$ 0.040 & 1.348 $\pm$ 0.113 & 1.281 $\pm$ 0.009 & 1.181 $\pm$ 0.005 & \textbf{1.125 $\pm$ 0.006} \\
          \cmidrule{2-8}
          & \multirow{2}[2]{*}{Book} & MAE & 1.027 $\pm$ 0.000 & 1.394 $\pm$ 0.040 & 1.237 $\pm$ 0.037 & 1.089 $\pm$ 0.007 & \textbf{0.952 $\pm$ 0.001} \\
          &       & RMSE & 1.306 $\pm$ 0.006 & 1.690 $\pm$ 0.060 & 1.537 $\pm$ 0.046 & 1.382 $\pm$ 0.005 & \textbf{1.212 $\pm$ 0.001} \\
          \cmidrule{2-8}
          & \multirow{2}[2]{*}{Elec} & MAE & 1.033 $\pm$ 0.040 & 1.270 $\pm$ 0.040 & 1.234 $\pm$ 0.014 & 1.062 $\pm$ 0.034 & \textbf{0.941 $\pm$ 0.011} \\
          &       & RMSE & 1.365 $\pm$ 0.001 & 1.571 $\pm$ 0.043 & 1.509 $\pm$ 0.015 & 1.357 $\pm$ 0.036 & \textbf{1.203 $\pm$ 0.006} \\
\midrule
    \multirow{8}[8]{*}{\rotatebox[origin=c]{90}{80\% Cold Start Rate}} & \multirow{2}[2]{*}{CD} & MAE & 1.059 $\pm$ 0.028 & 1.223 $\pm$ 0.018 & 1.153 $\pm$ 0.017 & 0.959 $\pm$ 0.007 & \textbf{0.850 $\pm$ 0.008} \\
          &       & RMSE & 1.392 $\pm$ 0.008 & 1.511 $\pm$ 0.021 & 1.456 $\pm$ 0.011 & 1.244 $\pm$ 0.002 & \textbf{1.126 $\pm$ 0.012} \\
          \cmidrule{2-8}
          & \multirow{2}[2]{*}{MV} & MAE & 0.987 $\pm$ 0.009 & 1.326 $\pm$ 0.019 & 1.263 $\pm$ 0.012 & 1.004 $\pm$ 0.005 & \textbf{0.907 $\pm$ 0.010} \\
          &       & RMSE & 1.289 $\pm$ 0.004 & 1.615 $\pm$ 0.015 & 1.558 $\pm$ 0.009 & 1.275 $\pm$ 0.007 & \textbf{1.174 $\pm$ 0.005} \\
          \cmidrule{2-8}
          & \multirow{2}[2]{*}{Book} & MAE & 1.053 $\pm$ 0.004 & 1.465 $\pm$ 0.018 & 1.243 $\pm$ 0.028 & 1.069 $\pm$ 0.008 & \textbf{0.952 $\pm$ 0.003} \\
          &       & RMSE & 1.358 $\pm$ 0.020 & 1.828 $\pm$ 0.033 & 1.545 $\pm$ 0.030 & 1.371 $\pm$ 0.007 & \textbf{1.224 $\pm$ 0.009} \\
          \cmidrule{2-8}
          & \multirow{2}[2]{*}{Elec} & MAE & 1.070 $\pm$ 0.001 & 1.486 $\pm$ 0.014 & 1.261 $\pm$ 0.014 & 1.148 $\pm$ 0.005 & \textbf{1.027 $\pm$ 0.003} \\
          &       & RMSE & 1.398 $\pm$ 0.005 & 1.804 $\pm$ 0.013 & 1.557 $\pm$ 0.016 & 1.430 $\pm$ 0.008 & \textbf{1.324 $\pm$ 0.008} \\
    \bottomrule
    \end{tabular}%
    }
  \label{ablation table}%
  \vspace{-0.8em}
\end{table}%

\subsection{Generalization Capacity of MARCO (RQ3)}

\begin{figure}[]
    \centering
    \includegraphics[width=0.97\linewidth]{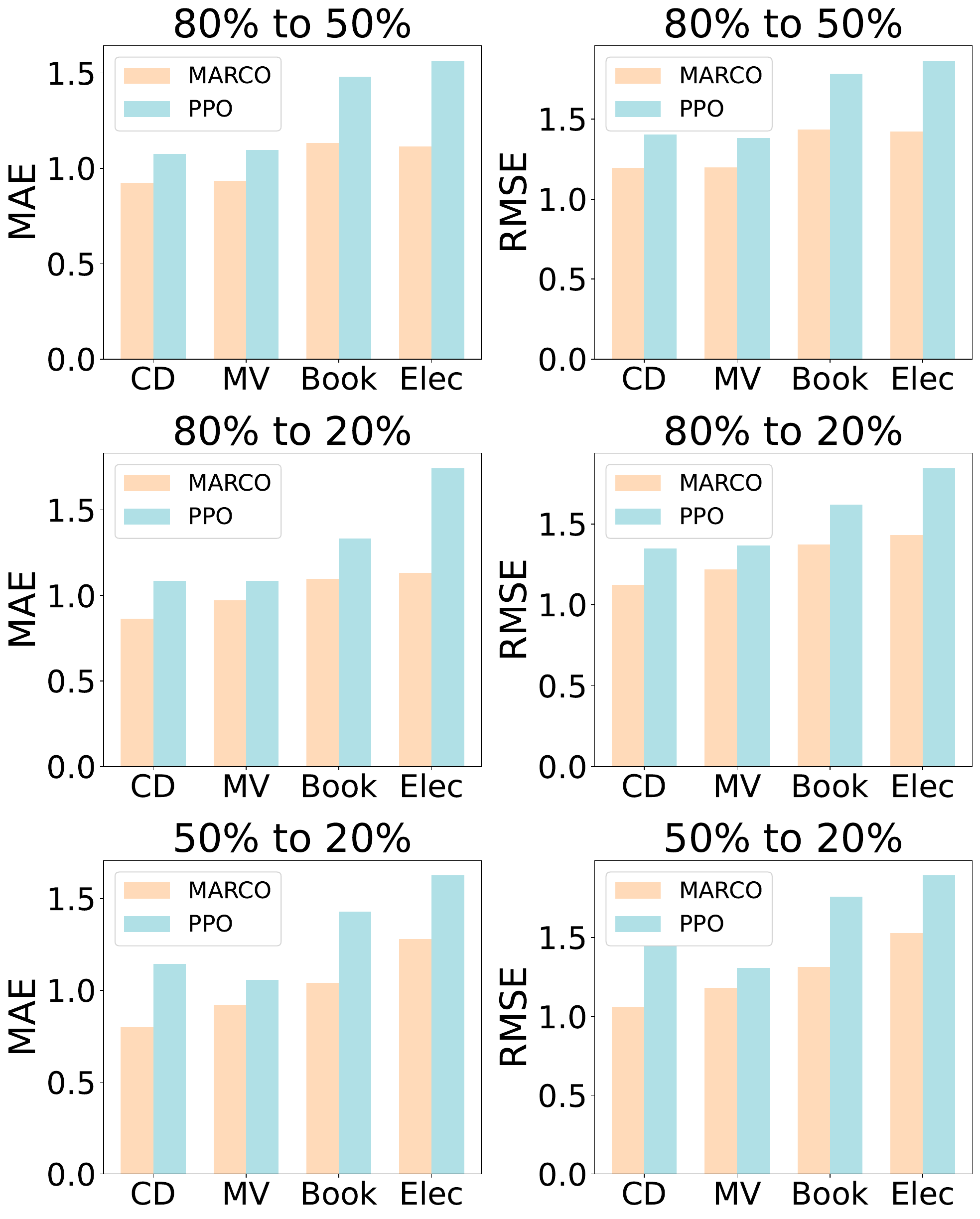} 
    \vspace{-0.5em}
    \caption{Transfer learning across cold start scenarios.}
    \label{fig3:transfer_fig}
\end{figure}

To verify the generalization ability of MARCO, which uses multi-agent reinforcement learning (MAPPO) compared to single-agent reinforcement learning (PPO) in transfer learning scenarios, experiments are conducted across four target domains with three transfer learning scenarios: (1) Models initially trained with a 50\% cold rate are applied to scenarios with a 20\% cold rate. (2) Models initially trained with an 80\% cold rate are applied to scenarios with a 20\% cold rate. (3) Models initially trained with an 80\% cold rate are applied to scenarios with a 50\% cold rate. Note that the model trained under one cold rate setting does not have access to users from another setting. This separation ensures the model's ability to perform transfer learning effectively. Figure~\ref{fig3:transfer_fig} presents a comparison of the results, demonstrating that the proposed framework (MARCO) consistently outperforms the single-agent framework (PPO) across all scenarios. By assigning each agent to each domain, MARCO learns the cooperative relationships among information from different source domains with heterogeneous distribution, minimizing the risk of overfitting and overestimation, while the single-agent method exhibits less robustness and poorer generalization.

\begin{table}[!t]
  \centering
  \caption{Effectiveness of entropy regularization.}
    \vspace{-0.5em}
    \resizebox{1\linewidth}{!}{
    \begin{tabular}{c|c|c|ccccc}
    \toprule
         & Target & Metric & PPO & PPO w/ Ent & MAPPO & MAPPO w/ Ent &  MARCO\\
          
\midrule 
    \multirow{8}[8]{*}{\rotatebox[origin=c]{90}{20\% Cold Start Rate}} & \multirow{2}[2]{*}{CD} & MAE & 1.088 $\pm$ 0.013 & 0.922 $\pm$ 0.006 & 0.936 $\pm$ 0.017 & 0.830 $\pm$ 0.003 & \textbf{0.823 $\pm$ 0.007} \\
          &       & RMSE & 1.361 $\pm$ 0.022 & 1.197 $\pm$ 0.005 & 1.195 $\pm$ 0.006 & {1.102 $\pm$ 0.002} & \textbf{1.087 $\pm$ 0.004} \\
          \cmidrule{2-8} 
          & \multirow{2}[2]{*}{MV} & MAE & 0.939 $\pm$ 0.009 & 0.936 $\pm$ 0.011 & 0.916 $\pm$ 0.004 & {0.934 $\pm$ 0.009} & \textbf{0.858 $\pm$ 0.004} \\
          &       & RMSE & 1.216 $\pm$ 0.006 & 1.193 $\pm$ 0.015 & 1.188 $\pm$ 0.003 & 1.193 $\pm$ 0.005 & \textbf{1.109 $\pm$ 0.002} \\
          \cmidrule{2-8} 
          & \multirow{2}[2]{*}{Book} & MAE & 1.118 $\pm$ 0.006 & 1.084 $\pm$ 0.001 & 1.056 $\pm$ 0.007 & {1.006 $\pm$ 0.041} & \textbf{0.925 $\pm$ 0.005} \\
          &       & RMSE & 1.403 $\pm$ 0.009 & 1.361 $\pm$ 0.005 & 1.354 $\pm$ 0.002 & {1.307 $\pm$ 0.038} & \textbf{1.171 $\pm$ 0.002} \\
          \cmidrule{2-8} 
          & \multirow{2}[2]{*}{Elec} & MAE & 1.055 $\pm$ 0.008 & 1.024 $\pm$ 0.010 & 1.001 $\pm$ 0.005 & {0.985 $\pm$ 0.005} & \textbf{0.980 $\pm$ 0.004} \\
          &       & RMSE & 1.354 $\pm$ 0.008 & 1.301 $\pm$ 0.008 & 1.296 $\pm$ 0.008 & {1.270 $\pm$ 0.003} & \textbf{1.258 $\pm$ 0.003} \\
\midrule
    \multirow{8}[8]{*}{\rotatebox[origin=c]{90}{50\% Cold Start Rate}} & \multirow{2}[2]{*}{CD} & MAE & 1.127 $\pm$ 0.012 & 0.975 $\pm$ 0.017 & 0.924 $\pm$ 0.009 & {0.835 $\pm$ 0.012} & \textbf{0.769 $\pm$ 0.011} \\
          &       & RMSE & 1.375 $\pm$ 0.011 & 1.272 $\pm$ 0.007 & 1.186 $\pm$ 0.011 & {1.107 $\pm$ 0.007} & \textbf{1.022 $\pm$ 0.005} \\
          \cmidrule{2-8}
          & \multirow{2}[2]{*}{MV} & MAE & 1.006 $\pm$ 0.005 & 0.957 $\pm$ 0.006 & 0.900 $\pm$ 0.006 & {0.892 $\pm$ 0.006} & \textbf{0.884 $\pm$ 0.008} \\
          &       & RMSE & 1.281 $\pm$ 0.009 & 1.239 $\pm$ 0.007 & 1.181 $\pm$ 0.005 & {1.154 $\pm$ 0.007} & \textbf{1.125 $\pm$ 0.006} \\
          \cmidrule{2-8}
          & \multirow{2}[2]{*}{Book} & MAE & 1.237 $\pm$ 0.037 & 1.197 $\pm$ 0.008 & 1.089 $\pm$ 0.007 & {0.968 $\pm$ 0.004} & \textbf{0.952 $\pm$ 0.001} \\
          &       & RMSE & 1.537 $\pm$ 0.046 & 1.480 $\pm$ 0.017 & 1.382 $\pm$ 0.005 & {1.245 $\pm$ 0.003} & \textbf{1.212 $\pm$ 0.001} \\
          \cmidrule{2-8}
          & \multirow{2}[2]{*}{Elec} & MAE & 1.234 $\pm$ 0.014 & 1.250 $\pm$ 0.006 & 1.062 $\pm$ 0.034 & {1.008 $\pm$ 0.044} & \textbf{0.941 $\pm$ 0.011} \\
          &       & RMSE & 1.509 $\pm$ 0.015 & 1.496 $\pm$ 0.010 & 1.357 $\pm$ 0.036 & {1.300 $\pm$ 0.038} & \textbf{1.203 $\pm$ 0.006} \\
\midrule
    \multirow{8}[8]{*}{\rotatebox[origin=c]{90}{80\% Cold Start Rate}} & \multirow{2}[2]{*}{CD} & MAE & 1.153 $\pm$ 0.017 & 1.108 $\pm$ 0.019 & 0.959 $\pm$ 0.007 & {0.852 $\pm$ 0.006} & \textbf{0.850 $\pm$ 0.008} \\
          &       & RMSE & 1.456 $\pm$ 0.011 & 1.375 $\pm$ 0.006 & 1.244 $\pm$ 0.002 & {1.134 $\pm$ 0.003} & \textbf{1.126 $\pm$ 0.012} \\
          \cmidrule{2-8}
          & \multirow{2}[2]{*}{MV} & MAE & 1.263 $\pm$ 0.012 & 1.076 $\pm$ 0.014 & 1.004 $\pm$ 0.005 & {0.923 $\pm$ 0.007} & \textbf{0.907 $\pm$ 0.010} \\
          &       & RMSE & 1.558 $\pm$ 0.009 & 1.348 $\pm$ 0.009 & 1.275 $\pm$ 0.007 & {1.204 $\pm$ 0.011} & \textbf{1.174 $\pm$ 0.005} \\
          \cmidrule{2-8}
          & \multirow{2}[2]{*}{Book} & MAE & 1.243 $\pm$ 0.028 & 1.130 $\pm$ 0.028 & 1.069 $\pm$ 0.008 & {1.022 $\pm$ 0.013} & \textbf{0.952 $\pm$ 0.003} \\
          &       & RMSE & 1.545 $\pm$ 0.030 & 1.421 $\pm$ 0.023 & 1.371 $\pm$ 0.007 & {1.304 $\pm$ 0.007} & \textbf{1.224 $\pm$ 0.009} \\
          \cmidrule{2-8}
          & \multirow{2}[2]{*}{Elec} & MAE & 1.261 $\pm$ 0.014 & 1.167 $\pm$ 0.020 & 1.148 $\pm$ 0.005 & {1.030 $\pm$ 0.005} & \textbf{1.027 $\pm$ 0.003} \\
          &       & RMSE & 1.557 $\pm$ 0.016 & 1.505 $\pm$ 0.008 & 1.430 $\pm$ 0.008 & {1.342 $\pm$ 0.005} & \textbf{1.324 $\pm$ 0.008} \\
    \bottomrule
    \end{tabular}%
    }
  \label{entropy table}%
\end{table}%

\subsection{The Robustness to Source Domains (RQ4)}
\begin{figure}[]
    \centering
    \includegraphics[width=0.97\linewidth]{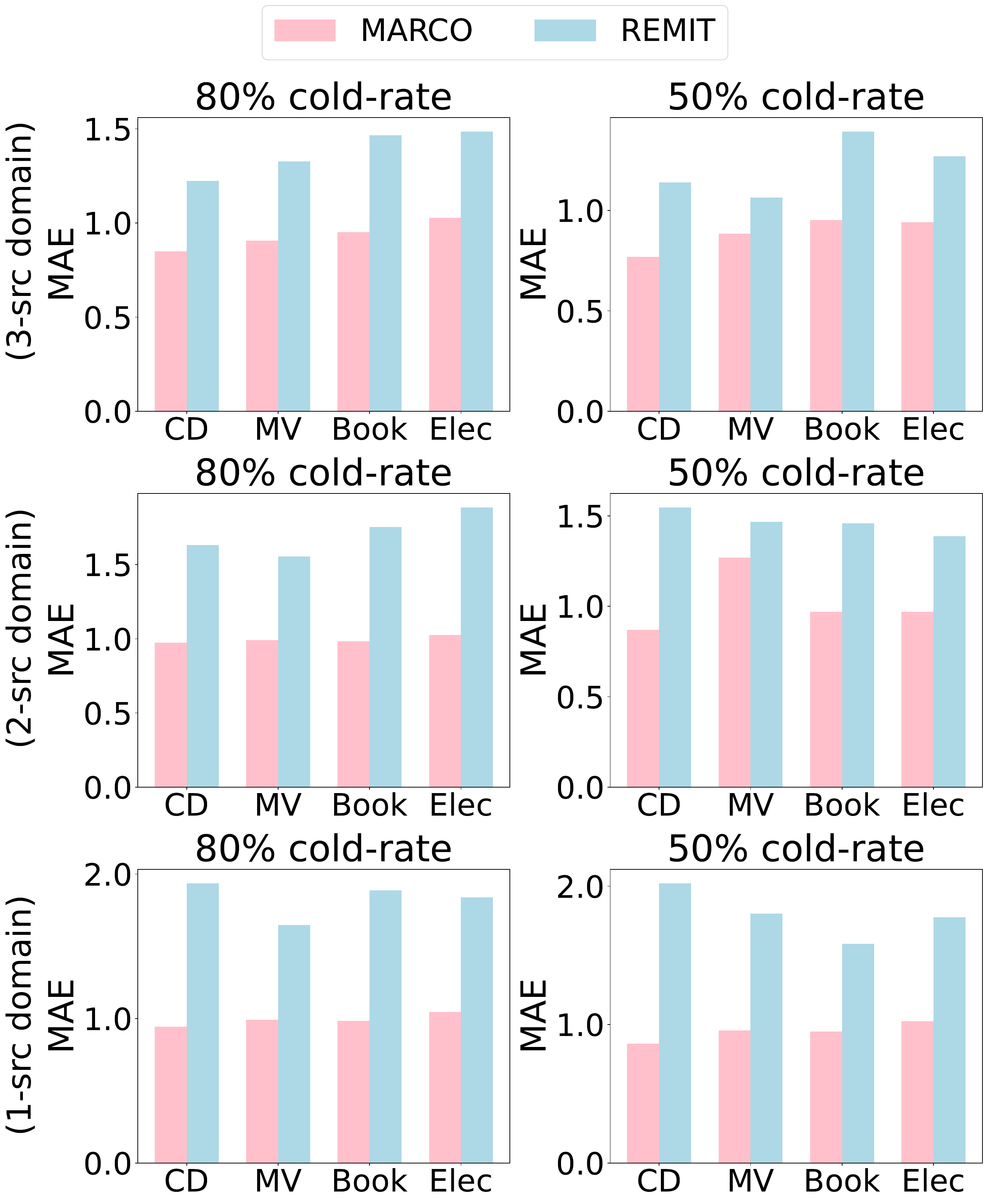} 
    \vspace{-0.5em}
    \caption{The robustness to the number of source domains }
    \label{agent_fig}
\end{figure}

To verify the robustness of MARCO compared to a single-agent reinforcement learning method (REMIT), the experiments under 50\% and 80\% cold-rate scenarios with varying numbers of source domains are conducted: (1) Models transfer knowledge from three source domains. (2) Models transfer knowledge from two source domains. (3) Models transfer knowledge from one source domain. The result detailed in Figure~\ref{agent_fig} demonstrates the superior robustness of the MARCO when the number of source domains is reduced. This is because the MARCO, by incorporating the cooperative nature with an entropy-based action diversity penalty, is able to promote diversity in joint agent behaviors, effectively capturing the heterogeneous characteristics of different source domains.

\subsection{Hyperparameter Sensitivity (RQ5)}
\begin{figure}[]
    \centering
    \includegraphics[width=0.97\linewidth]{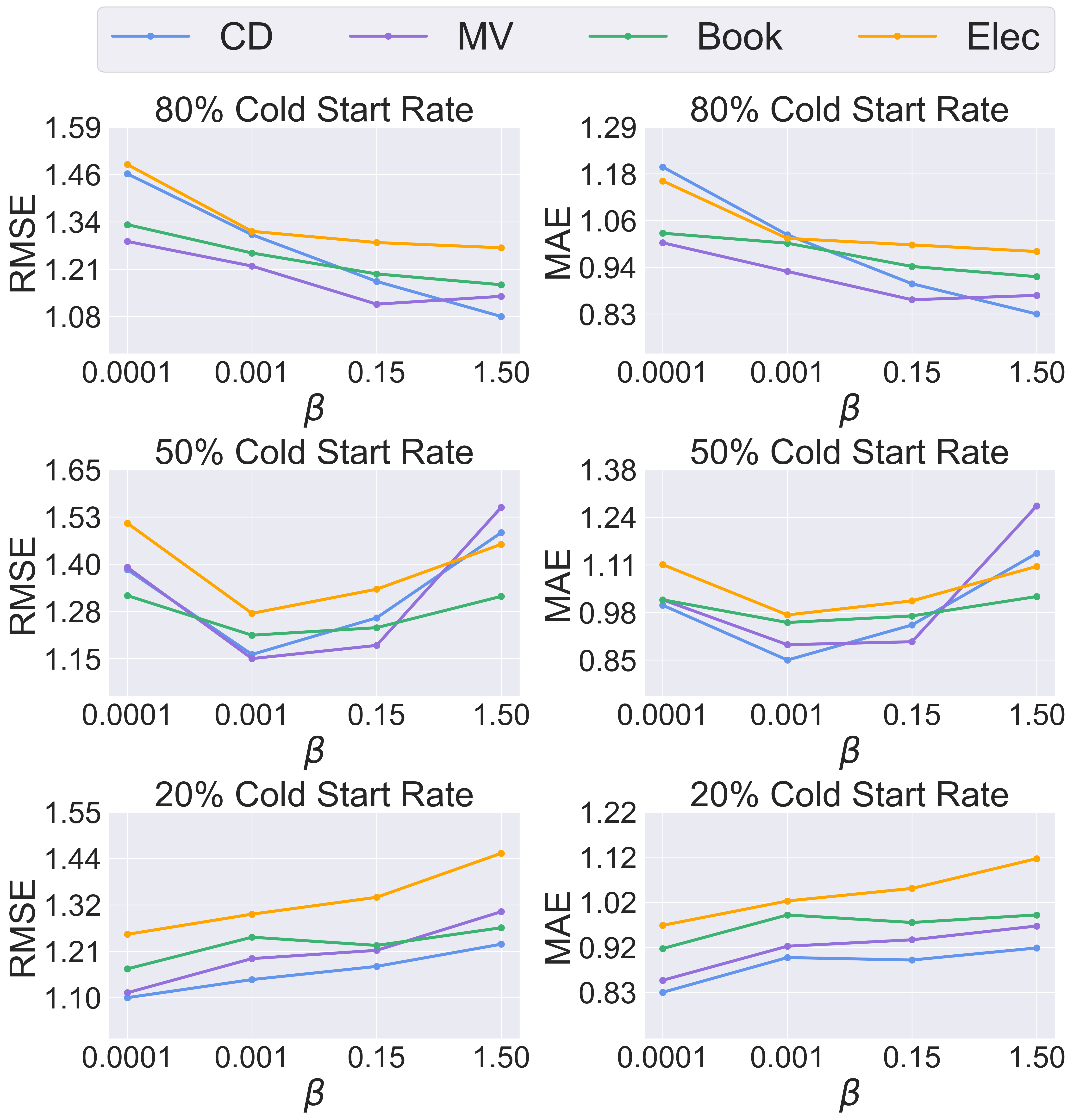} 
    \caption{Parameter sensitivity of $\beta$}
    \label{para_fig}
    \vspace{-2em}
\end{figure}

In this experiment, the sensitivity of the hyperparameter $\beta$, which is the coefficient of the entropy term in the objective function of MARCO in Eq.~(\ref{entropy_loss}), is investigated. The experiment detailed in Figure~\ref{para_fig} shows the relationship between $\beta$ and performance in different cold start scenarios. When the cold start rate is 20\%, setting $\beta$ to 0.0001 achieves the best performance across four domains because the model can learn valuable information from abundant data with minimal regularization to prevent overfitting. As the cold start rate increases, the optimal $\beta$ values also increase, set to 0.001 for a 50\% cold start rate and 1.5 for an 80\% cold start rate. The results indicate that entropy regularization indeed benefits performance.

\section{Conclusion and Future Work}
In this paper, MARCO, a cooperative multi-agent reinforcement learning-based framework, is presented. It is specifically designed for multi-source cross-domain recommendation (CDR). MARCO effectively addresses the critical challenges of negative transfer by assigning dedicated agents to estimate contributions from multiple source domains individually, ensuring efficient credit assignment and optimized knowledge transfer. By integrating an entropy-based action diversity penalty, MARCO enhances policy expressiveness and stabilizes training, enabling the framework to robustly manage the heterogeneity and distributional discrepancies inherent in cross-domain recommendation scenarios. Extensive empirical evaluations conducted across four benchmark datasets confirm MARCO's superiority over state-of-the-art CDR methods, demonstrating significant improvements in recommendation accuracy, robustness, and generalization capabilities. 

Future work will focus on extending MARCO’s architecture to integrate richer domain-specific signals and further optimizing the multi-agent cooperation mechanism to improve scalability and robustness in more complex and heterogeneous recommendation environments. Moreover, evaluating MARCO across a wider spectrum of cross-domain datasets and investigating its integration with diverse pre-training strategies to understand their impact on CDR performance remain promising directions.

\section{Acknowledegments}
This work is supported by projects DE200101610, DE250100919, CE200100025 funded by Australian Research Council, and CSIRO’s Science Leader Project R-91559.

\bibliographystyle{ACM-Reference-Format}
\bibliography{MARCO}


\begin{thebibliography}{72}


\ifx \showCODEN    \undefined \def \showCODEN     #1{\unskip}     \fi
\ifx \showISBNx    \undefined \def \showISBNx     #1{\unskip}     \fi
\ifx \showISBNxiii \undefined \def \showISBNxiii  #1{\unskip}     \fi
\ifx \showISSN     \undefined \def \showISSN      #1{\unskip}     \fi
\ifx \showLCCN     \undefined \def \showLCCN      #1{\unskip}     \fi
\ifx \shownote     \undefined \def \shownote      #1{#1}          \fi
\ifx \showarticletitle \undefined \def \showarticletitle #1{#1}   \fi
\ifx \showURL      \undefined \def \showURL       {\relax}        \fi
\providecommand\bibfield[2]{#2}
\providecommand\bibinfo[2]{#2}
\providecommand\natexlab[1]{#1}
\providecommand\showeprint[2][]{arXiv:#2}

\bibitem[Chang et~al\mbox{.}(2023)]%
        {pep}
\bibfield{author}{\bibinfo{person}{Jianxin Chang}, \bibinfo{person}{Chenbin Zhang}, \bibinfo{person}{Yiqun Hui}, \bibinfo{person}{Dewei Leng}, \bibinfo{person}{Yanan Niu}, \bibinfo{person}{Yang Song}, {and} \bibinfo{person}{Kun Gai}.} \bibinfo{year}{2023}\natexlab{}.
\newblock \showarticletitle{Pepnet: Parameter and embedding personalized network for infusing with personalized prior information}. In \bibinfo{booktitle}{\emph{SIGKDD}}.
\newblock


\bibitem[Chen et~al\mbox{.}(2018)]%
        {q-stab}
\bibfield{author}{\bibinfo{person}{Shi-Yong Chen}, \bibinfo{person}{Yang Yu}, \bibinfo{person}{Qing Da}, \bibinfo{person}{Jun Tan}, \bibinfo{person}{Hai-Kuan Huang}, {and} \bibinfo{person}{Hai-Hong Tang}.} \bibinfo{year}{2018}\natexlab{}.
\newblock \showarticletitle{Stabilizing reinforcement learning in dynamic environment with application to online recommendation}. In \bibinfo{booktitle}{\emph{KDD}}.
\newblock


\bibitem[Chen et~al\mbox{.}(2013)]%
        {ma_re_mul}
\bibfield{author}{\bibinfo{person}{Wei Chen}, \bibinfo{person}{Wynne Hsu}, {and} \bibinfo{person}{Mong{-}Li Lee}.} \bibinfo{year}{2013}\natexlab{}.
\newblock \showarticletitle{Making recommendations from multiple domains}. In \bibinfo{booktitle}{\emph{SIGKDD}}.
\newblock


\bibitem[Cui et~al\mbox{.}(2020)]%
        {HeroGRAPH}
\bibfield{author}{\bibinfo{person}{Qiang Cui}, \bibinfo{person}{Tao Wei}, \bibinfo{person}{Yafeng Zhang}, {and} \bibinfo{person}{Qing Zhang}.} \bibinfo{year}{2020}\natexlab{}.
\newblock \showarticletitle{HeroGRAPH: A Heterogeneous Graph Framework for Multi-Target Cross-Domain Recommendation}.
\newblock \bibinfo{journal}{\emph{ORSUM@ RecSys}} (\bibinfo{year}{2020}).
\newblock


\bibitem[Fu et~al\mbox{.}(2019)]%
        {addition-singebridge_fusingreviews}
\bibfield{author}{\bibinfo{person}{Wenjing Fu}, \bibinfo{person}{Zhaohui Peng}, \bibinfo{person}{Senzhang Wang}, \bibinfo{person}{Yang Xu}, {and} \bibinfo{person}{Jin Li}.} \bibinfo{year}{2019}\natexlab{}.
\newblock \showarticletitle{Deeply Fusing Reviews and Contents for Cold Start Users in Cross-Domain Recommendation Systems}. In \bibinfo{booktitle}{\emph{AAAI}}.
\newblock


\bibitem[Gui et~al\mbox{.}(2019)]%
        {marl-twitter}
\bibfield{author}{\bibinfo{person}{Tao Gui}, \bibinfo{person}{Peng Liu}, \bibinfo{person}{Qi Zhang}, \bibinfo{person}{Liang Zhu}, \bibinfo{person}{Minlong Peng}, \bibinfo{person}{Yunhua Zhou}, {and} \bibinfo{person}{Xuanjing Huang}.} \bibinfo{year}{2019}\natexlab{}.
\newblock \showarticletitle{Mention recommendation in Twitter with cooperative multi-agent reinforcement learning}. In \bibinfo{booktitle}{\emph{SIGIR}}.
\newblock


\bibitem[He et~al\mbox{.}(2018)]%
        {multi-cf}
\bibfield{author}{\bibinfo{person}{Ming He}, \bibinfo{person}{Jiuling Zhang}, \bibinfo{person}{Peng Yang}, {and} \bibinfo{person}{Kaisheng Yao}.} \bibinfo{year}{2018}\natexlab{}.
\newblock \showarticletitle{Robust transfer learning for cross-domain collaborative filtering using multiple rating patterns approximation}. In \bibinfo{booktitle}{\emph{WSDM}}.
\newblock


\bibitem[He et~al\mbox{.}(2020)]%
        {marl-withoutcomu}
\bibfield{author}{\bibinfo{person}{Xu He}, \bibinfo{person}{Bo An}, \bibinfo{person}{Yanghua Li}, \bibinfo{person}{Haikai Chen}, \bibinfo{person}{Rundong Wang}, \bibinfo{person}{Xinrun Wang}, \bibinfo{person}{Runsheng Yu}, \bibinfo{person}{Xin Li}, {and} \bibinfo{person}{Zhirong Wang}.} \bibinfo{year}{2020}\natexlab{}.
\newblock \showarticletitle{Learning to collaborate in multi-module recommendation via multi-agent reinforcement learning without communication}. In \bibinfo{booktitle}{\emph{RecSys}}.
\newblock


\bibitem[He et~al\mbox{.}(2017)]%
        {ncf}
\bibfield{author}{\bibinfo{person}{Xiangnan He}, \bibinfo{person}{Lizi Liao}, \bibinfo{person}{Hanwang Zhang}, \bibinfo{person}{Liqiang Nie}, \bibinfo{person}{Xia Hu}, {and} \bibinfo{person}{Tat-Seng Chua}.} \bibinfo{year}{2017}\natexlab{}.
\newblock \showarticletitle{Neural collaborative filtering}. In \bibinfo{booktitle}{\emph{WWW}}.
\newblock


\bibitem[Hu et~al\mbox{.}(2018)]%
        {MTNET}
\bibfield{author}{\bibinfo{person}{Guangneng Hu}, \bibinfo{person}{Yu Zhang}, {and} \bibinfo{person}{Qiang Yang}.} \bibinfo{year}{2018}\natexlab{}.
\newblock \showarticletitle{MTNet: a neural approach for cross-domain recommendation with unstructured text}.
\newblock \bibinfo{journal}{\emph{KDD Deep Learning Day}} (\bibinfo{year}{2018}).
\newblock


\bibitem[Hu et~al\mbox{.}(2013a)]%
        {addition-mulcf_cft}
\bibfield{author}{\bibinfo{person}{Liang Hu}, \bibinfo{person}{Jian Cao}, \bibinfo{person}{Guandong Xu}, \bibinfo{person}{Longbing Cao}, \bibinfo{person}{Zhiping Gu}, {and} \bibinfo{person}{Can Zhu}.} \bibinfo{year}{2013}\natexlab{a}.
\newblock \showarticletitle{Personalized recommendation via cross-domain triadic factorization}. In \bibinfo{booktitle}{\emph{WWW}}.
\newblock


\bibitem[Hu et~al\mbox{.}(2013b)]%
        {addition-mulcf_tf}
\bibfield{author}{\bibinfo{person}{Liang Hu}, \bibinfo{person}{Jian Cao}, \bibinfo{person}{Guandong Xu}, \bibinfo{person}{Longbing Cao}, \bibinfo{person}{Zhiping Gu}, {and} \bibinfo{person}{Can Zhu}.} \bibinfo{year}{2013}\natexlab{b}.
\newblock \showarticletitle{Personalized recommendation via cross-domain triadic factorization}. In \bibinfo{booktitle}{\emph{WWW}}.
\newblock


\bibitem[ki~Leung et~al\mbox{.}(2007)]%
        {clare}
\bibfield{author}{\bibinfo{person}{Cane~Wing ki Leung}, \bibinfo{person}{Stephen~Chi fai Chan}, {and} \bibinfo{person}{Fu lai Chung}.} \bibinfo{year}{2007}\natexlab{}.
\newblock \showarticletitle{Applying cross-level association rule mining to cold-start recommendations}. In \bibinfo{booktitle}{\emph{Wi-IAT}}.
\newblock


\bibitem[Kim et~al\mbox{.}(2019)]%
        {multi-rnn}
\bibfield{author}{\bibinfo{person}{Donghyun Kim}, \bibinfo{person}{Sungchul Kim}, \bibinfo{person}{Handong Zhao}, \bibinfo{person}{Sheng Li}, \bibinfo{person}{Ryan~A. Rossi}, {and} \bibinfo{person}{Eunyee Koh}.} \bibinfo{year}{2019}\natexlab{}.
\newblock \showarticletitle{Domain switch-aware holistic recurrent neural network for modeling multi-domain user behavior}. In \bibinfo{booktitle}{\emph{WSDM}}.
\newblock


\bibitem[Kingma and Ba(2014)]%
        {adam}
\bibfield{author}{\bibinfo{person}{Diederik~P Kingma} {and} \bibinfo{person}{Jimmy Ba}.} \bibinfo{year}{2014}\natexlab{}.
\newblock \showarticletitle{Adam: A method for stochastic optimization}.
\newblock \bibinfo{journal}{\emph{arXiv preprint}} (\bibinfo{year}{2014}).
\newblock


\bibitem[Kuhnle et~al\mbox{.}(2021)]%
        {reinforce_retrival}
\bibfield{author}{\bibinfo{person}{Alexander Kuhnle}, \bibinfo{person}{Miguel Aroca-Ouellette}, \bibinfo{person}{Anindya Basu}, \bibinfo{person}{Murat Sensoy}, \bibinfo{person}{John Reid}, {and} \bibinfo{person}{Dell Zhang}.} \bibinfo{year}{2021}\natexlab{}.
\newblock \showarticletitle{Reinforcement learning for information retrieval}. In \bibinfo{booktitle}{\emph{SIGIR}}.
\newblock


\bibitem[Li et~al\mbox{.}(2023a)]%
        {allforone}
\bibfield{author}{\bibinfo{person}{Chenglin Li}, \bibinfo{person}{Yuanzhen Xie}, \bibinfo{person}{Chenyun Yu}, \bibinfo{person}{Bo Hu}, \bibinfo{person}{Zang Li}, \bibinfo{person}{Guoqiang Shu}, \bibinfo{person}{Xiaohu Qie}, {and} \bibinfo{person}{Di Niu}.} \bibinfo{year}{2023}\natexlab{a}.
\newblock \showarticletitle{One for all, all for one: Learning and transferring user embeddings for cross-domain recommendation}. In \bibinfo{booktitle}{\emph{WSDM}}.
\newblock


\bibitem[Li et~al\mbox{.}(2017)]%
        {attention-net}
\bibfield{author}{\bibinfo{person}{Jing Li}, \bibinfo{person}{Pengjie Ren}, \bibinfo{person}{Zhumin Chen}, \bibinfo{person}{Zhaochun Ren}, \bibinfo{person}{Tao Lian}, {and} \bibinfo{person}{Jun Ma}.} \bibinfo{year}{2017}\natexlab{}.
\newblock \showarticletitle{Neural attentive session-based recommendation}. In \bibinfo{booktitle}{\emph{CIKM}}.
\newblock


\bibitem[Li et~al\mbox{.}(2019)]%
        {cd-fm}
\bibfield{author}{\bibinfo{person}{Lile Li}, \bibinfo{person}{Quan Do}, {and} \bibinfo{person}{Wei Liu}.} \bibinfo{year}{2019}\natexlab{}.
\newblock \showarticletitle{Cross-domain recommendation via coupled factorization machines}. In \bibinfo{booktitle}{\emph{AAAI}}.
\newblock


\bibitem[Li et~al\mbox{.}(2023b)]%
        {ham}
\bibfield{author}{\bibinfo{person}{Xiaopeng Li}, \bibinfo{person}{Fan Yan}, \bibinfo{person}{Xiangyu Zhao}, \bibinfo{person}{Yichao Wang}, \bibinfo{person}{Bo Chen}, \bibinfo{person}{Huifeng Guo}, {and} \bibinfo{person}{Ruiming Tang}.} \bibinfo{year}{2023}\natexlab{b}.
\newblock \showarticletitle{Hamur: Hyper adapter for multi-domain recommendation}. In \bibinfo{booktitle}{\emph{CIKM}}.
\newblock


\bibitem[Li et~al\mbox{.}(2021)]%
        {li2021discovering}
\bibfield{author}{\bibinfo{person}{Yang Li}, \bibinfo{person}{Tong Chen}, \bibinfo{person}{Yadan Luo}, \bibinfo{person}{Hongzhi Yin}, {and} \bibinfo{person}{Zi Huang}.} \bibinfo{year}{2021}\natexlab{}.
\newblock \showarticletitle{Discovering Collaborative Signals for Next POI Recommendation with Iterative Seq2Graph Augmentation}. In \bibinfo{booktitle}{\emph{IJCAI}}.
\newblock


\bibitem[Liu et~al\mbox{.}(2020a)]%
        {DDPG-end2end}
\bibfield{author}{\bibinfo{person}{Feng Liu}, \bibinfo{person}{Huifeng Guo}, \bibinfo{person}{Xutao Li}, \bibinfo{person}{Ruiming Tang}, \bibinfo{person}{Yunming Ye}, {and} \bibinfo{person}{Xiuqiang He}.} \bibinfo{year}{2020}\natexlab{a}.
\newblock \showarticletitle{End-to-end deep reinforcement learning based recommendation with supervised embedding}. In \bibinfo{booktitle}{\emph{WSDM}}.
\newblock


\bibitem[Liu et~al\mbox{.}(2020b)]%
        {DDPG-Topaware}
\bibfield{author}{\bibinfo{person}{Feng Liu}, \bibinfo{person}{Ruiming Tang}, \bibinfo{person}{Huifeng Guo}, \bibinfo{person}{Xutao Li}, \bibinfo{person}{Yunming Ye}, {and} \bibinfo{person}{Xiuqiang He}.} \bibinfo{year}{2020}\natexlab{b}.
\newblock \showarticletitle{Top-aware reinforcement learning based recommendation}.
\newblock \bibinfo{journal}{\emph{Neurocomputing}} (\bibinfo{year}{2020}).
\newblock


\bibitem[Liu et~al\mbox{.}(2018)]%
        {DDPG-DRR}
\bibfield{author}{\bibinfo{person}{Feng Liu}, \bibinfo{person}{Ruiming Tang}, \bibinfo{person}{Xutao Li}, \bibinfo{person}{Weinan Zhang}, \bibinfo{person}{Yunming Ye}, \bibinfo{person}{Haokun Chen}, \bibinfo{person}{Huifeng Guo}, {and} \bibinfo{person}{Yuzhou Zhang}.} \bibinfo{year}{2018}\natexlab{}.
\newblock \showarticletitle{Deep reinforcement learning based recommendation with explicit user-item interactions modeling}.
\newblock \bibinfo{journal}{\emph{arXiv preprint}} (\bibinfo{year}{2018}).
\newblock


\bibitem[Liu et~al\mbox{.}(2024)]%
        {graph-reviewer}
\bibfield{author}{\bibinfo{person}{Jing Liu}, \bibinfo{person}{Lele Sun}, \bibinfo{person}{Weizhi Nie}, \bibinfo{person}{Peiguang Jing}, {and} \bibinfo{person}{Yuting Su}.} \bibinfo{year}{2024}\natexlab{}.
\newblock \showarticletitle{Graph disentangled contrastive learning with personalized transfer for cross-domain recommendation}. In \bibinfo{booktitle}{\emph{AAAI}}.
\newblock


\bibitem[Liu et~al\mbox{.}(2021)]%
        {graph-single1}
\bibfield{author}{\bibinfo{person}{Ziqi Liu}, \bibinfo{person}{Yue Shen}, \bibinfo{person}{Xiaocheng Cheng}, \bibinfo{person}{Qiang Li}, \bibinfo{person}{Jianping Wei}, \bibinfo{person}{Zhiqiang Zhang}, \bibinfo{person}{Dong Wang}, \bibinfo{person}{Xiaodong Zeng}, \bibinfo{person}{Jinjie Gu}, {and} \bibinfo{person}{Jun Zhou}.} \bibinfo{year}{2021}\natexlab{}.
\newblock \showarticletitle{Learning representations of inactive users: A cross domain approach with graph neural networks}. In \bibinfo{booktitle}{\emph{CIKM}}.
\newblock


\bibitem[Loni et~al\mbox{.}(2014)]%
        {addition-mulcf_cfm}
\bibfield{author}{\bibinfo{person}{Babak Loni}, \bibinfo{person}{Yue Shi}, \bibinfo{person}{Martha Larso}, {and} \bibinfo{person}{Alan Hanjalic}.} \bibinfo{year}{2014}\natexlab{}.
\newblock \showarticletitle{Cross-domain collaborative filtering with factorization machines}. In \bibinfo{booktitle}{\emph{ECIR}}.
\newblock


\bibitem[Ma et~al\mbox{.}(2019)]%
        {pinet}
\bibfield{author}{\bibinfo{person}{Muyang Ma}, \bibinfo{person}{Pengjie Ren}, \bibinfo{person}{Yujie Lin}, \bibinfo{person}{Zhumin Chen}, \bibinfo{person}{Jun Ma}, {and} \bibinfo{person}{Maarten{-}de Rijke}.} \bibinfo{year}{2019}\natexlab{}.
\newblock \showarticletitle{$\pi$-net: A parallel information-sharing network for shared-account cross-domain sequential recommendations}. In \bibinfo{booktitle}{\emph{SIGIR}}.
\newblock


\bibitem[Man et~al\mbox{.}(2017)]%
        {emb_cdr}
\bibfield{author}{\bibinfo{person}{Tong Man}, \bibinfo{person}{Huawei Shen}, \bibinfo{person}{Xiaolong Jin}, {and} \bibinfo{person}{Xueqi Cheng}.} \bibinfo{year}{2017}\natexlab{}.
\newblock \showarticletitle{Cross-Domain Recommendation: An Embedding and Mapping Approach}. In \bibinfo{booktitle}{\emph{IJCAI}}.
\newblock


\bibitem[Nguyen et~al\mbox{.}(2018)]%
        {nguyen2018credit}
\bibfield{author}{\bibinfo{person}{Duc~Thien Nguyen}, \bibinfo{person}{Akshat Kumar}, {and} \bibinfo{person}{Hoong~Chuin Lau}.} \bibinfo{year}{2018}\natexlab{}.
\newblock \showarticletitle{Credit assignment for collective multiagent RL with global rewards}.
\newblock \bibinfo{journal}{\emph{NIPS}} (\bibinfo{year}{2018}).
\newblock


\bibitem[Ouyang et~al\mbox{.}(2019)]%
        {graph-multisource}
\bibfield{author}{\bibinfo{person}{Yi Ouyang}, \bibinfo{person}{Bin Guo}, \bibinfo{person}{Xing Tang}, \bibinfo{person}{Xiuqiang He}, \bibinfo{person}{Jian Xiong}, {and} \bibinfo{person}{Zhiwen Yu}.} \bibinfo{year}{2019}\natexlab{}.
\newblock \showarticletitle{Learning cross-domain representation with multi-graph neural network}.
\newblock \bibinfo{journal}{\emph{arXiv preprint arXiv:1905.10095}} (\bibinfo{year}{2019}).
\newblock


\bibitem[Ouyang et~al\mbox{.}(2021)]%
        {graph-multisource2}
\bibfield{author}{\bibinfo{person}{Yi Ouyang}, \bibinfo{person}{Bin Guo}, \bibinfo{person}{Xing Tang}, \bibinfo{person}{Xiuqiang He}, \bibinfo{person}{Jian Xiong}, {and} \bibinfo{person}{Zhiwen Yu}.} \bibinfo{year}{2021}\natexlab{}.
\newblock \showarticletitle{Mobile app cross-domain recommendation with multi-graph neural network}.
\newblock \bibinfo{journal}{\emph{ACM Transactions on Knowledge Discovery from Data (TKDD)}} (\bibinfo{year}{2021}).
\newblock


\bibitem[Pan et~al\mbox{.}(2019)]%
        {pg-con}
\bibfield{author}{\bibinfo{person}{Feiyang Pan}, \bibinfo{person}{Qingpeng Cai}, \bibinfo{person}{Pingzhong Tang}, \bibinfo{person}{Fuzhen Zhuang}, {and} \bibinfo{person}{Qing He}.} \bibinfo{year}{2019}\natexlab{}.
\newblock \showarticletitle{Policy gradients for contextual recommendations}. In \bibinfo{booktitle}{\emph{WWW}}.
\newblock


\bibitem[Pan et~al\mbox{.}(2010)]%
        {addition-singcf_spar}
\bibfield{author}{\bibinfo{person}{Weike Pan}, \bibinfo{person}{Evan Xiang}, \bibinfo{person}{Nathan Liu}, {and} \bibinfo{person}{Qiang Yang}.} \bibinfo{year}{2010}\natexlab{}.
\newblock \showarticletitle{Transfer learning in collaborative filtering for sparsity reduction}. In \bibinfo{booktitle}{\emph{AAAI}}.
\newblock


\bibitem[Pan and Yang(2013)]%
        {addition-singcf-trans}
\bibfield{author}{\bibinfo{person}{Weike Pan} {and} \bibinfo{person}{Qiang Yang}.} \bibinfo{year}{2013}\natexlab{}.
\newblock \showarticletitle{Transfer learning in heterogeneous collaborative filtering domains}.
\newblock \bibinfo{journal}{\emph{Artificial intelligence}} (\bibinfo{year}{2013}).
\newblock


\bibitem[Petruzzelli et~al\mbox{.}(2024)]%
        {llm-instructing}
\bibfield{author}{\bibinfo{person}{Alessandro Petruzzelli}, \bibinfo{person}{Cataldo Musto}, \bibinfo{person}{Lucrezia Laraspata}, \bibinfo{person}{Ivan Rinaldi}, \bibinfo{person}{Marco de Gemmis}, \bibinfo{person}{Pasquale Lops}, {and} \bibinfo{person}{Giovanni Semeraro}.} \bibinfo{year}{2024}\natexlab{}.
\newblock \showarticletitle{Instructing and prompting large language models for explainable cross-domain recommendations}. In \bibinfo{booktitle}{\emph{RecSys}}.
\newblock


\bibitem[Qiu et~al\mbox{.}(2021)]%
        {qiu2021memory}
\bibfield{author}{\bibinfo{person}{Ruihong Qiu}, \bibinfo{person}{Zi Huang}, {and} \bibinfo{person}{Hongzhi Yin}.} \bibinfo{year}{2021}\natexlab{}.
\newblock \showarticletitle{Memory augmented multi-instance contrastive predictive coding for sequential recommendation}. In \bibinfo{booktitle}{\emph{ICDM}}.
\newblock


\bibitem[Qiu et~al\mbox{.}(2022a)]%
        {ntk}
\bibfield{author}{\bibinfo{person}{Ruihong Qiu}, \bibinfo{person}{Zi Huang}, {and} \bibinfo{person}{Hongzhi Yin}.} \bibinfo{year}{2022}\natexlab{a}.
\newblock \showarticletitle{Beyond Double Ascent via Recurrent Neural Tangent Kernel in Sequential Recommendation}. In \bibinfo{booktitle}{\emph{ICDM}}.
\newblock


\bibitem[Qiu et~al\mbox{.}(2022b)]%
        {qiu2022contrastive}
\bibfield{author}{\bibinfo{person}{Ruihong Qiu}, \bibinfo{person}{Zi Huang}, \bibinfo{person}{Hongzhi Yin}, {and} \bibinfo{person}{Zijian Wang}.} \bibinfo{year}{2022}\natexlab{b}.
\newblock \showarticletitle{Contrastive learning for representation degeneration problem in sequential recommendation}. In \bibinfo{booktitle}{\emph{WSDM}}.
\newblock


\bibitem[Qiu et~al\mbox{.}(2019)]%
        {qiu2019rethinking}
\bibfield{author}{\bibinfo{person}{Ruihong Qiu}, \bibinfo{person}{Jingjing Li}, \bibinfo{person}{Zi Huang}, {and} \bibinfo{person}{Hongzhi Yin}.} \bibinfo{year}{2019}\natexlab{}.
\newblock \showarticletitle{Rethinking the item order in session-based recommendation with graph neural networks}. In \bibinfo{booktitle}{\emph{CIKM}}.
\newblock


\bibitem[Qiu et~al\mbox{.}(2020)]%
        {qiu2020gag}
\bibfield{author}{\bibinfo{person}{Ruihong Qiu}, \bibinfo{person}{Hongzhi Yin}, \bibinfo{person}{Zi Huang}, {and} \bibinfo{person}{Tong Chen}.} \bibinfo{year}{2020}\natexlab{}.
\newblock \showarticletitle{Gag: Global attributed graph neural network for streaming session-based recommendation}. In \bibinfo{booktitle}{\emph{SIGIR}}.
\newblock


\bibitem[Rashid et~al\mbox{.}(2020)]%
        {QMIX}
\bibfield{author}{\bibinfo{person}{Tabish Rashid}, \bibinfo{person}{Mikayel Samvelyan}, \bibinfo{person}{Christian~Schroeder De~Witt}, \bibinfo{person}{Gregory Farquhar}, \bibinfo{person}{Jakob Foerster}, {and} \bibinfo{person}{Shimon Whiteson}.} \bibinfo{year}{2020}\natexlab{}.
\newblock \showarticletitle{Monotonic value function factorisation for deep multi-agent reinforcement learning}.
\newblock \bibinfo{journal}{\emph{Journal of Machine Learning Research}} (\bibinfo{year}{2020}).
\newblock


\bibitem[Rendle et~al\mbox{.}(2012)]%
        {bpr}
\bibfield{author}{\bibinfo{person}{Steffen Rendle}, \bibinfo{person}{Christoph Freudenthaler}, \bibinfo{person}{Zeno Gantner}, {and} \bibinfo{person}{Lars Schmidt-Thieme}.} \bibinfo{year}{2012}\natexlab{}.
\newblock \showarticletitle{BPR: Bayesian personalized ranking from implicit feedback}.
\newblock \bibinfo{journal}{\emph{arXiv preprint}} (\bibinfo{year}{2012}).
\newblock


\bibitem[Schulman et~al\mbox{.}(2017)]%
        {PPO}
\bibfield{author}{\bibinfo{person}{John Schulman}, \bibinfo{person}{Filip Wolski}, \bibinfo{person}{Prafulla Dhariwal}, \bibinfo{person}{Alec Radford}, {and} \bibinfo{person}{Oleg Klimov}.} \bibinfo{year}{2017}\natexlab{}.
\newblock \showarticletitle{Proximal policy optimization algorithms}.
\newblock \bibinfo{journal}{\emph{arXiv preprint}} (\bibinfo{year}{2017}).
\newblock


\bibitem[Singh et~al\mbox{.}(2008)]%
        {CMF}
\bibfield{author}{\bibinfo{person}{Ajit{-}P Singh}, \bibinfo{person}{Geoffrey{-}J Gordon}, {and} \bibinfo{person}{Emre Sargin}.} \bibinfo{year}{2008}\natexlab{}.
\newblock \showarticletitle{Relational learning via collective matrix factorization}. In \bibinfo{booktitle}{\emph{SIGKDD}}.
\newblock


\bibitem[Sun et~al\mbox{.}(2023)]%
        {remit}
\bibfield{author}{\bibinfo{person}{Caiqi Sun}, \bibinfo{person}{Jiewei Gu}, \bibinfo{person}{BinBin Hu}, \bibinfo{person}{Xin Dong}, \bibinfo{person}{Hai Li}, \bibinfo{person}{Lei Cheng}, {and} \bibinfo{person}{Linjian Mo}.} \bibinfo{year}{2023}\natexlab{}.
\newblock \showarticletitle{REMIT: reinforced multi-interest transfer for cross-domain recommendation}. In \bibinfo{booktitle}{\emph{AAAI}}.
\newblock


\bibitem[Wang et~al\mbox{.}(2021)]%
        {graph-single2}
\bibfield{author}{\bibinfo{person}{Chen Wang}, \bibinfo{person}{Yueqing Liang}, \bibinfo{person}{Zhiwei Liu}, \bibinfo{person}{Tao Zhang}, {and} \bibinfo{person}{S.~Yu Philip}.} \bibinfo{year}{2021}\natexlab{}.
\newblock \showarticletitle{Pre-training graph neural network for cross domain recommendation}. In \bibinfo{booktitle}{\emph{CogMI}}.
\newblock


\bibitem[Wang et~al\mbox{.}(2024)]%
        {marl_recom}
\bibfield{author}{\bibinfo{person}{Zhefan Wang}, \bibinfo{person}{Yuanqing Yu}, \bibinfo{person}{Wendi Zheng}, \bibinfo{person}{Weizhi Ma}, {and} \bibinfo{person}{Min Zhang}.} \bibinfo{year}{2024}\natexlab{}.
\newblock \showarticletitle{Multi-Agent Collaboration Framework for Recommender Systems}. In \bibinfo{booktitle}{\emph{arXiv preprint}}.
\newblock


\bibitem[Williams(1992)]%
        {reinforce}
\bibfield{author}{\bibinfo{person}{Ronald~J. Williams}.} \bibinfo{year}{1992}\natexlab{}.
\newblock \showarticletitle{Simple statistical gradient-following algorithms for connectionist reinforcement learning}.
\newblock \bibinfo{journal}{\emph{Machine Learning}} (\bibinfo{year}{1992}).
\newblock


\bibitem[Wu et~al\mbox{.}(2022)]%
        {dynamic_rl_retrival}
\bibfield{author}{\bibinfo{person}{Junda Wu}, \bibinfo{person}{Zhihui Xie}, \bibinfo{person}{Tong Yu}, \bibinfo{person}{Handong Zhao}, \bibinfo{person}{Ruiyi Zhang}, {and} \bibinfo{person}{Shuai Li}.} \bibinfo{year}{2022}\natexlab{}.
\newblock \showarticletitle{Dynamics-aware adaptation for reinforcement learning based cross-domain interactive recommendation}. In \bibinfo{booktitle}{\emph{SIGIR}}.
\newblock


\bibitem[Xie et~al\mbox{.}(2021)]%
        {DDPG-hier}
\bibfield{author}{\bibinfo{person}{Ruobing Xie}, \bibinfo{person}{Shaoliang Zhang}, \bibinfo{person}{Rui Wang}, \bibinfo{person}{Feng Xia}, {and} \bibinfo{person}{Leyu Lin}.} \bibinfo{year}{2021}\natexlab{}.
\newblock \showarticletitle{Hierarchical reinforcement learning for integrated recommendation}. In \bibinfo{booktitle}{\emph{AAAI}}.
\newblock


\bibitem[Yang et~al\mbox{.}(2017)]%
        {multi-mpf}
\bibfield{author}{\bibinfo{person}{Chunfeng Yang}, \bibinfo{person}{Huan Yan}, \bibinfo{person}{Donghan Yu}, \bibinfo{person}{Yong Li}, {and} \bibinfo{person}{Dah~Ming Chiu}.} \bibinfo{year}{2017}\natexlab{}.
\newblock \showarticletitle{Multi-site User Behavior Modeling and Its Application in Video Recommendation}. In \bibinfo{booktitle}{\emph{SIGIR}}.
\newblock


\bibitem[Ye et~al\mbox{.}(2023)]%
        {data-distribution-sparsity}
\bibfield{author}{\bibinfo{person}{Xiaoxin Ye}, \bibinfo{person}{Yun Li}, {and} \bibinfo{person}{Lina Yao}.} \bibinfo{year}{2023}\natexlab{}.
\newblock \showarticletitle{DREAM: Decoupled representation via extraction attention module and supervised contrastive learning for cross-domain sequential recommender}. In \bibinfo{booktitle}{\emph{RecSys}}.
\newblock


\bibitem[Yu et~al\mbox{.}(2022)]%
        {mappo}
\bibfield{author}{\bibinfo{person}{Chao Yu}, \bibinfo{person}{Akash Velu}, \bibinfo{person}{Eugene Vinitsky}, \bibinfo{person}{Jiaxuan Gao}, \bibinfo{person}{Yu Wang}, \bibinfo{person}{Alexandre Bayen}, {and} \bibinfo{person}{Yi Wu}.} \bibinfo{year}{2022}\natexlab{}.
\newblock \showarticletitle{The surprising effectiveness of ppo in cooperative multi-agent games}.
\newblock \bibinfo{journal}{\emph{Advances in Neural Information Processing Systems}} (\bibinfo{year}{2022}).
\newblock


\bibitem[Yu et~al\mbox{.}(2020)]%
        {semi-cf}
\bibfield{author}{\bibinfo{person}{Wenhui Yu}, \bibinfo{person}{Xiao Lin}, \bibinfo{person}{Junfeng Ge}, \bibinfo{person}{Wenwu Ou}, {and} \bibinfo{person}{Zheng Qin}.} \bibinfo{year}{2020}\natexlab{}.
\newblock \showarticletitle{Semi-supervised collaborative filtering by text-enhanced domain adaptation}. In \bibinfo{booktitle}{\emph{SIGKDD}}.
\newblock


\bibitem[Yuan et~al\mbox{.}(2020)]%
        {multi-parashare}
\bibfield{author}{\bibinfo{person}{Fajie Yuan}, \bibinfo{person}{Xiangnan He}, \bibinfo{person}{Alexandros Karatzoglou}, {and} \bibinfo{person}{Liguang Zhang}.} \bibinfo{year}{2020}\natexlab{}.
\newblock \showarticletitle{Parameter-efficient transfer from sequential behaviors for user modeling and recommendation}. In \bibinfo{booktitle}{\emph{SIGIR}}.
\newblock


\bibitem[Zhang et~al\mbox{.}(2019)]%
        {pg-text}
\bibfield{author}{\bibinfo{person}{Ruiyi Zhang}, \bibinfo{person}{Tong Yu}, \bibinfo{person}{Yilin Shen}, \bibinfo{person}{Hongxia Jin}, \bibinfo{person}{Changyou Chen}, {and} \bibinfo{person}{Lawrence Carin}.} \bibinfo{year}{2019}\natexlab{}.
\newblock \showarticletitle{Text-based interactive recommendation via constraint-augmented reinforcement learning}. In \bibinfo{booktitle}{\emph{NeurIPS}}.
\newblock


\bibitem[Zhang et~al\mbox{.}(2022)]%
        {negative_transfer}
\bibfield{author}{\bibinfo{person}{Wen Zhang}, \bibinfo{person}{Lingfei Deng}, \bibinfo{person}{Lei Zhang}, {and} \bibinfo{person}{Dongrui Wu}.} \bibinfo{year}{2022}\natexlab{}.
\newblock \showarticletitle{A survey on negative transfer}.
\newblock \bibinfo{journal}{\emph{IEEE/CAA Journal of Automatica Sinica}} (\bibinfo{year}{2022}).
\newblock


\bibitem[Zhang et~al\mbox{.}(2021)]%
        {marl-vech}
\bibfield{author}{\bibinfo{person}{Weijia Zhang}, \bibinfo{person}{Hao Liu}, \bibinfo{person}{Fan Wang}, \bibinfo{person}{Tong Xu}, \bibinfo{person}{Haoran Xin}, \bibinfo{person}{Dejing Dou}, {and} \bibinfo{person}{Hui Xiong}.} \bibinfo{year}{2021}\natexlab{}.
\newblock \showarticletitle{Intelligent electric vehicle charging recommendation based on multi-agent reinforcement learning}. In \bibinfo{booktitle}{\emph{WWW}}.
\newblock


\bibitem[Zhang et~al\mbox{.}(2024)]%
        {yi2024roler}
\bibfield{author}{\bibinfo{person}{Yi Zhang}, \bibinfo{person}{Ruihong Qiu}, \bibinfo{person}{Jiajun Liu}, {and} \bibinfo{person}{Sen Wang}.} \bibinfo{year}{2024}\natexlab{}.
\newblock \showarticletitle{ROLeR: Effective Reward Shaping in Offline Reinforcement Learning for Recommender Systems}. In \bibinfo{booktitle}{\emph{CIKM}}.
\newblock


\bibitem[Zhang et~al\mbox{.}(2025)]%
        {yi2025darlr}
\bibfield{author}{\bibinfo{person}{Yi Zhang}, \bibinfo{person}{Ruihong Qiu}, \bibinfo{person}{Xuwei Xu}, \bibinfo{person}{Jiajun Liu}, {and} \bibinfo{person}{Sen Wang}.} \bibinfo{year}{2025}\natexlab{}.
\newblock \showarticletitle{DARLR: Dual-Agent Offline Reinforcement Learning for Recommender Systems with Dynamic Reward}. In \bibinfo{booktitle}{\emph{SIGIR}}.
\newblock


\bibitem[Zhao et~al\mbox{.}(2020a)]%
        {CATN}
\bibfield{author}{\bibinfo{person}{Cheng Zhao}, \bibinfo{person}{Chenliang Li}, \bibinfo{person}{Rong Xiao}, \bibinfo{person}{Hongbo Deng}, {and} \bibinfo{person}{Aixin Sun}.} \bibinfo{year}{2020}\natexlab{a}.
\newblock \showarticletitle{CATN: Cross-domain recommendation for cold-start users via aspect transfer network}. In \bibinfo{booktitle}{\emph{SIGIR}}.
\newblock


\bibitem[Zhao et~al\mbox{.}(2020b)]%
        {marl-Mahrl}
\bibfield{author}{\bibinfo{person}{Dongyang Zhao}, \bibinfo{person}{Liang Zhang}, \bibinfo{person}{Bo Zhang}, \bibinfo{person}{Lizhou Zheng}, \bibinfo{person}{Yongjun Bao}, {and} \bibinfo{person}{Weipeng Yan}.} \bibinfo{year}{2020}\natexlab{b}.
\newblock \showarticletitle{Mahrl: Multi-goals abstraction based deep hierarchical reinforcement learning for recommendations}. In \bibinfo{booktitle}{\emph{SIGIR}}.
\newblock


\bibitem[Zhao et~al\mbox{.}(2022)]%
        {multisparse}
\bibfield{author}{\bibinfo{person}{Xiaoyun Zhao}, \bibinfo{person}{Ning Yang}, {and} \bibinfo{person}{Philip{-}S Yu}.} \bibinfo{year}{2022}\natexlab{}.
\newblock \showarticletitle{Multi-sparse-domain collaborative recommendation via enhanced comprehensive aspect preference learning}. In \bibinfo{booktitle}{\emph{WSDM}}.
\newblock


\bibitem[Zhao et~al\mbox{.}(2019)]%
        {DDPG-list}
\bibfield{author}{\bibinfo{person}{Xiangyu Zhao}, \bibinfo{person}{Liang Zhang}, \bibinfo{person}{Long Xia}, \bibinfo{person}{Zhuoye Ding}, \bibinfo{person}{Dawei Yin}, {and} \bibinfo{person}{Jiliang Tang}.} \bibinfo{year}{2019}\natexlab{}.
\newblock \showarticletitle{Deep reinforcement learning for list-wise recommendations}. In \bibinfo{booktitle}{\emph{DRL4KDD}}.
\newblock


\bibitem[Zhao et~al\mbox{.}(2020c)]%
        {marl-joint}
\bibfield{author}{\bibinfo{person}{Xiangyu Zhao}, \bibinfo{person}{Xudong Zheng}, \bibinfo{person}{Xiwang Yang}, \bibinfo{person}{Xiaobing Liu}, {and} \bibinfo{person}{Jiliang Tang}.} \bibinfo{year}{2020}\natexlab{c}.
\newblock \showarticletitle{Jointly learning to recommend and advertise}. In \bibinfo{booktitle}{\emph{SIGKDD}}.
\newblock


\bibitem[Zheng et~al\mbox{.}(2018)]%
        {q-drn}
\bibfield{author}{\bibinfo{person}{Guanjie Zheng}, \bibinfo{person}{Fuzheng Zhang}, \bibinfo{person}{Zihan Zheng}, \bibinfo{person}{Yang Xiang}, \bibinfo{person}{Nicholas~Jing Yuan}, \bibinfo{person}{Xing Xie}, {and} \bibinfo{person}{Zhenhui Li}.} \bibinfo{year}{2018}\natexlab{}.
\newblock \showarticletitle{DRN: A deep reinforcement learning framework for news recommendation}. In \bibinfo{booktitle}{\emph{WWW}}.
\newblock


\bibitem[Zhu et~al\mbox{.}(2017)]%
        {cdr_survey}
\bibfield{author}{\bibinfo{person}{Feng Zhu}, \bibinfo{person}{Yan Wang}, \bibinfo{person}{Chaochao Chen}, \bibinfo{person}{Jun Zhou}, \bibinfo{person}{Longfei}, {and} \bibinfo{person}{Guanfeng Liu}.} \bibinfo{year}{2017}\natexlab{}.
\newblock \showarticletitle{Cross-Domain Recommendation: Challenges, Progress, and Prospects}. In \bibinfo{booktitle}{\emph{IJCAI}}.
\newblock


\bibitem[Zhu et~al\mbox{.}(2021)]%
        {addition-mulbri-gam}
\bibfield{author}{\bibinfo{person}{Feng Zhu}, \bibinfo{person}{Yan Wang}, \bibinfo{person}{Jun Zhou}, \bibinfo{person}{Chaochao Chen}, \bibinfo{person}{Longfei Li}, {and} \bibinfo{person}{Guanfeng Liu}.} \bibinfo{year}{2021}\natexlab{}.
\newblock \showarticletitle{A unified framework for cross-domain and cross-system recommendations}.
\newblock \bibinfo{journal}{\emph{IEEE Transactions on Knowledge and Data Engineering}} (\bibinfo{year}{2021}).
\newblock


\bibitem[Zhu et~al\mbox{.}(2023)]%
        {graph-cdr}
\bibfield{author}{\bibinfo{person}{Jiajie Zhu}, \bibinfo{person}{Yan Wang}, \bibinfo{person}{Feng Zhu}, {and} \bibinfo{person}{Zhu Sun}.} \bibinfo{year}{2023}\natexlab{}.
\newblock \showarticletitle{Domain disentanglement with interpolative data augmentation for dual-target cross-domain recommendation}. In \bibinfo{booktitle}{\emph{RecSys}}.
\newblock


\bibitem[Zhu et~al\mbox{.}(2022a)]%
        {personalized—transfer-revier}
\bibfield{author}{\bibinfo{person}{Yongchun Zhu}, \bibinfo{person}{Zhenwei Tang}, \bibinfo{person}{Yudan Liu}, \bibinfo{person}{Fuzhen Zhuang}, \bibinfo{person}{Ruobing Xie}, \bibinfo{person}{Xu Zhang}, \bibinfo{person}{Leyu Lin}, {and} \bibinfo{person}{Qing He}.} \bibinfo{year}{2022}\natexlab{a}.
\newblock \showarticletitle{Personalized transfer of user preferences for cross-domain recommendation}. In \bibinfo{booktitle}{\emph{WSDM}}.
\newblock


\bibitem[Zhu et~al\mbox{.}(2022b)]%
        {ptup}
\bibfield{author}{\bibinfo{person}{Yongchun Zhu}, \bibinfo{person}{Zhenwei Tang}, \bibinfo{person}{Yudan Liu}, \bibinfo{person}{Fuzhen Zhuang}, \bibinfo{person}{Ruobing Xie}, \bibinfo{person}{Xu Zhang}, \bibinfo{person}{Leyu Lin}, {and} \bibinfo{person}{Qing He}.} \bibinfo{year}{2022}\natexlab{b}.
\newblock \showarticletitle{Personalized transfer of user preferences for cross-domain recommendation}. In \bibinfo{booktitle}{\emph{WSDM}}.
\newblock


\end{thebibliography}

\end{document}